\documentclass[a4paper,fleqn,usenatbib]{mnras}

\usepackage{newtxtext,newtxmath}

\usepackage{graphicx}   
\usepackage{amsmath}    
\usepackage{amssymb}    
\usepackage{natbib}

\title[Extended $\ion{He}{ii}$ emission from the Slug nebula]{The large and small scale properties of the intergalactic gas in the Slug Ly$\alpha$ nebula revealed by MUSE $\ion{He}{ii}$ emission observations} 

\author[S. Cantalupo et al.]{Sebastiano Cantalupo$^{1}$\thanks{E-mail: cantalupo@phys.ethz.ch},
Gabriele Pezzulli$^{1}$,
Simon J. Lilly$^{1}$,
Raffaella Anna Marino$^{1}$,
\newauthor
Sofia G. Gallego$^{1}$,
Joop Schaye$^{2}$,
Roland Bacon$^{3}$,
Anna Feltre$^{3}$,
Wolfram Kollatschny$^{4}$,
\newauthor
Themiya Nanayakkara$^{2}$,
Johan Richard$^{3}$,
Martin Wendt$^{5,6}$,
Lutz Wisotzki$^{6}$ and
\newauthor
J. Xavier Prochaska$^{7}$
\\
$^{1}$ Department of Physics, ETH Zurich, Wolfgang-Pauli-Strasse 27, 8093, Zurich, Switzerland\\
$^{2}$ Leiden Observatory, Leiden University, PO Box 9513, 2300 RA Leiden, The Netherlands.\\
$^{3}$ Univ Lyon, Univ Lyon1, Ens de Lyon, CNRS, Centre de Recherche Astrophysique de Lyon UMR5574, F-69230, Saint-Genis-Laval, France. \\
$^{4}$ Institut f\"ur Astrophysik, Universität G\"ottingen, Friedrich-Hund-Platz 1, D-37077 G\"ottingen, Germany.\\
$^{5}$ Institut f\"ur Physik und Astronomie, Karl-Liebknecht-Str. 24/25, D-14476 Potsdam/Golm, Germany.\\
$^{6}$ Leibniz-Institut für Astrophysik Potsdam (AIP), An der Sternwarte 16, D-14482 Potsdam, Germany.\\
$^{7}$ UCO/Lick Observatory, 1156 High St., UC Santa Cruz, Santa Cruz, CA 95064.\\
}

\date{Accepted 2018 December 19. Received 2018 December 19. In original form 2018 July 19.}

\pubyear{2018}

\begin{document}
\label{firstpage}
\pagerange{\pageref{firstpage}--\pageref{lastpage}}
\maketitle
\begin{abstract}
With a projected size of about 450 kpc at z$\simeq2.3$, the Slug Ly$\alpha$ nebula is a rare laboratory to study,
in emission, the properties of the intergalactic gas in the Cosmic Web. Since its discovery, the
Slug has been the subject of several spectroscopic follow-ups to constrain the properties of the emitting gas. 
Here we report the results of a deep MUSE integral-field spectroscopic search for non-resonant, extended $\ion{He}{ii}\lambda1640$  and metal emission. 
Extended $\ion{He}{ii}$ radiation is detected on scales of about 100 kpc, but only in some regions
associated with the bright Ly$\alpha$ emission 
 {and a continuum-detected source},
 implying large and abrupt variations in the line ratios across adjacent regions in projected space. 
The recent detection of associated H$\alpha$ emission 
 {and similar abrupt variations in the Ly$\alpha$ kinematics,}
strongly suggest that the $\ion{He}{ii}$/Ly$\alpha$ gradient is due to large variations
in the physical distances between the associated quasar and these regions.
This implies that the overall length of the emitting structure could extend to  {physical} Mpc scales and 
be mostly oriented along our line of sight. 
At the same time, the relatively low $\ion{He}{ii}$/Ly$\alpha$ values suggest that the emitting gas 
has a broad density distribution that - if expressed in terms of a lognormal - implies dispersions as high as 
those expected in the interstellar medium of galaxies. 
These results strengthen the possibility that the density distribution of intergalactic gas at high-redshift 
is extremely clumpy and multiphase on scales below our current  {observational} spatial resolution of a few  {physical} kpc. 
\end{abstract}

\begin{keywords}
galaxies:haloes -- galaxies: high-redshift -- intergalactic medium -- quasars: emission lines -- cosmology: observations. 
\end{keywords}

\section{Introduction}

Our standard cosmological paradigm predicts that both 
dark and baryonic matter in the universe
should be distributed in a network of filaments that we call
the Cosmic Web where galaxies form and evolve
\citep[e.g.,][]{bond96}. 
During the last few years, a new observational window on the
densest part of this Cosmic Web has been opened 
by the direct detection of hydrogen in Ly$\alpha$ emission
on large intergalactic
\begin{footnote}{
in this work, we will use the term ``intergalactic" in his broadest
sense, i.e. including the material that is in proximity of galaxies
(but outside of the interstellar medium) and typically 
indicated as ``circumgalactic medium" in the recent literature.
}\end{footnote}
 scales in proximity of 
bright quasars 
(e.g., \citealt{sc14,martin14,hennawi15,borisova16}; see also \citealt{sc17} for a review).
These two-dimensional (or three-dimensional in the case of integral-field spectroscopy) 
observations with spatial resolution currently limited only by the atmospheric seeing 
(corresponding to a few kpc at z$\sim3$) are now complementing several decades 
of absorption studies (see e.g., \citealt{rauch98,meiksin09} for reviews)
albeit still on possibly different environments.
The latter are limited to either one-dimensional
or very sparse two-dimensional probes of the Intergalactic Medium (IGM) 
with spatial resolution of a few Mpc \citep[e.g.][]{lee17}. 

The possibility of detecting the IGM in emission by using, e.g., fluorescent Ly$\alpha$
due to the cosmic UV background, was already suggested several decades ago 
\citep[e.g.][]{hogan87,gould96}. However, the faintness of the expected emission 
is hampering the possibility of detecting such emission with current facilities 
(e.g., \citealt{rauch08}; see also \citealt{sc05} and \citealt{gallego18} for discussion).
By looking around 
bright quasars, 
the expected fluorescent emission due to recombination radiation should be boosted 
by several orders  of magnitude within the densest part of the cosmic web
\citep[e.g.][]{sc05,kollmeier10,sc12}.
Deep narrow-band imaging
campaigns around bright quasars \citep[e.g.][]{sc12, sc14, martin14, hennawi15, fab16}
and, more recently, integral-field  spectroscopic campaigns with 
the Multi Unit Spectroscopic Explorer (MUSE) (e.g., \citealt{borisova16,fab18}) 
and the Palomar/Keck Cosmic Web Imager (P/KCWI) (e.g., \citealt{martin14,cai18})
are finally revealing giant Ly$\alpha$ nebulae with size exceeding 100 kpc 
around essentially all bright quasars (at least at $z>3$ 
 {while at $z\simeq2$ they seem to be detected more rarely,} 
see e.g., \citealt{fab16} and \citealt{sc17} for discussion).

The Slug nebula at z$\simeq$2.3 was one of the first, largest and most luminous among
the nebulae found in these observations \citep{sc14}. It is characterised by very bright 
and filamentary Ly$\alpha$ emission extending about 450 projected kpc around 
the quasar UM287 (see Fig.\ref{FigRGB}). 
As discussed in \citet{sc14} and \citet{sc17}, the
high Ly$\alpha$ Surface Brightness (SB) of the Slug would imply either: 
i) very large densities 
of cold ($T\sim10^{4}$ K) and ionised gas (if emission is dominated by hydrogen recombinations) 
or, ii) very large column
densities of neutral hydrogen (if the emission is due to ''photon-pumping'' or scattering
of Ly$\alpha$ photons produced within the quasar broad line region). 

Unfortunately, Ly$\alpha$ imaging alone does not help disentangle
these two emission mechanisms. Several spectroscopic follow-ups by means
of long-slit observations have tried recently to detect other non-resonant lines such as 
$\ion{He}{ii}\lambda1640$ (i.e., the first line of the Balmer series of singly ionized helium; see \citealt{fab15}) 
and hydrogen H$\alpha$ \citep{leibler18} in the Slug. 
At the same time, the Slug Ly$\alpha$ emission has been re-observed with integral-field
spectroscopy using the Palomar Cosmic Web Imager (PCWI) by \citet{martin15}, revealing large velocity 
shifts that, at the limited spatial resolution of PCWI have been interpreted
as a possible signature of a rotating structure on 100 kpc scales. Given the resonant nature
of the Ly$\alpha$ emission, it is not clear however how much of these shifts are due to radiative
transfer effects rather than kinematics. A two-dimensional velocity map of a non-resonant line 
would be essential to understand the possible kinematical signatures
in the nebula. However, until now only long-slit detection or upper limits on H$\alpha$ or $\ion{He}{ii}$1640
are available for some parts of the nebula, as discussed below.

The non-detection of $\ion{He}{ii}\lambda1640$ in a low-resolution LRIS long-slit spectroscopic 
observation of \citet{fab15} resulted in a $\ion{He}{ii}$/Ly$\alpha$ upper limit of 0.18 ($3\sigma$) 
in the brightest part of the nebula, suggesting either that Ly$\alpha$ emission is produced 
by ''photon-pumping'' (the second scenario in \citealt{sc14}) or, e.g., that the ionisation parameter 
in some part of the nebula is relatively small (log($U)<-1.5$; 
 {see \citet{fab15} for discussion} 
)
. Assuming a ''single-density scenario''
(or a ``delta-function'' density distribution as discussed in this work)
where cold gas is in the form of clumps, a single distance of 160 kpc
and a plausible flux from UM287 this
upper limit on U would translate into a gas density of $n>3$ cm$^{-3}$. 
However, such non-detection does not give us a constraint on the
emission mechanism and is obviously limited to the small region
covered by the spectroscopic slit. 

By means of long-slit IR spectroscopy with MOSFIRE of part of the Slug,
\citet{leibler18} were able to detect H$\alpha$ with a flux similar to the expected
recombination radiation scenario for Ly$\alpha$.
This result clearly rules out that, at least in the region covered by the slit,
''photon-pumping'' has a significant contribution to the Ly$\alpha$ emission.
In this case, deep $\ion{He}{ii}$1640 constraints can be used to infer gas densities
with some assumptions about the quasar ionizing flux.

At the same time the relatively narrow H$\alpha$ emission 
{(with a velocity dispersion of about 180 km/s)}, 
compared to the Ly$\alpha$ line width in similar regions
 {(showing a velocity dispersion of about 400 km/s)}, 
does suggest the presence of radiative transfer effects.
The medium-resolution Ly$\alpha$ spectrum obtained by \citet{leibler18} using LRIS in the same
study shows however a similar velocity centroid for Ly$\alpha$ and the integrated
H$\alpha$, suggesting that Ly$\alpha$ could still be used as a tracer of kinematics,
at least in a average sense and on large scales. Different from the PCWI spectrum,
the Ly$\alpha$ velocity shifts seem very abrupt on spatially adjacent regions, hinting to
the possibility of more complex kinematics than the simple rotating structure 
suggested by \citet{martin15} or that the Slug could be composed of different systems
separated in velocity (and possibly physical) space. 

In this study, we use MUSE (see e.g., \citealt{bacon15}) to overcome some of the major limitations 
of long-slit spectroscopic observations discussed above in order to obtain full two-dimensional 
and kinematic constraints on the non-resonant $\ion{He}{ii}\lambda1640$ emission and metal lines
at high spatial resolution (seeing-limited). Combined with previous studies 
of both Ly$\alpha$ and H$\alpha$ emission, our
deep integral field observations allow us to address several open questions, 
including: i) what is the density distribution of the cold gas in the
intergalactic medium around the Slug quasar (UM287)
on scales below a few kpc?, ii) what are the large-scale properties and the
kinematics of the intergalactic filament(s) associated with the Slug nebula?

Before addressing these questions, we will go through a description of our
experimental design, observations, data reduction and analysis in section \ref{sec_obs}.
We will then present our main results in section \ref{sec_res} followed by a discussion
of how our results address the questions above in section \ref{Disc}. 
We will then summarise our work in section \ref{Sum}.
Throughout the paper we use the cosmological parameters: 
$h=0.696$, $\Omega_{\rm{m}}=0.286$, and $\Omega_{\rm{vac}}=0.714$
as derived by \citet{bennett14}. Angular size distances have been computed
using \citet{wright06} providing a scale of 8.371 kpc/'' at $z=2.279$.
Distances are always proper, unless stated otherwise.

\section{Observations and data reduction}\label{sec_obs}

The field of the Slug nebula (quasar UM287; \citealt{sc14}) was observed with MUSE
during two visitor-mode runs in P94 as a part of the MUSE Guaranteed Time
of Observations (GTO) program (proposal ID: 094.A-0396) for a total of 9 hours
of exposure time on source. Data acquisition followed the standard strategy for
our GTO programs on quasar fields: 36 individual exposures of 15 minutes
integration time each were taken applying a small dithering 
and rotation of 90 degrees between them
(see also \citealt{borisova16,marino18}). 
Nights were classified as clear with a median
seeing of about 0.8"
 {as obtained from the measurement of the quasar Point Spread Function (PSF).}
The only available configuration in P94
(and subsequent periods until P100) was the Wide Field Mode without Adaptive Optics 
(WFM-NAO) providing a field of view of about $1\times1$ arcmin$^2$
sampled by 90000 spaxels with spatial sizes of $0.2\times0.2$ arcsec$^2$
and spectral resolution elements with sizes of $1.25\mathrm{\AA}$. 
 We chose the nominal wavelength mode resulting
in a wavelength coverage extending from 4750$\mathrm{\AA}$ to 9350$\mathrm{\AA}$.

At the measured systemic redshift of the Slug quasar, UM287, i.e.  {$z=2.283\pm0.001$ 
obtained from the detection of a narrow (FWHM$=200$ km/s) and compact 
CO(3-2) emission line (De Carli et al, in prep.)}, the wavelength
coverage in the rest-frame extends from about 1447$\mathrm{\AA}$ to
 about 2848$\mathrm{\AA}$. 
 This allows us to cover the expected brightest
 UV emission lines after Ly$\alpha$ such as the $\ion{C}{iv}\lambda1549$ doublet 
 (5081.3-5089.8$\mathrm{\AA}$ in the observed frame in air),
 $\ion{He}{ii}\lambda1640$ (5384.0$\mathrm{\AA}$ in the observed frame in air),
 and the $\ion{C}{iii}\lambda1908$ doublet (6257.9-6264.6$\mathrm{\AA}$ in the observed frame in air).
 The MgII2796 doublet is in principle also covered by our observations
 although we expect this line to appear at the very red edge of our wavelength range where
 the instrumental sensitivity, instrumental systematics and bright sky lines significantly reduce
 our ability to put constraints on this line, as discussed in section \ref{Disc}.
 
Data reduction followed a combination of both standard recipes from the MUSE
pipeline (version 1.6, \citealt{DRS}) and custom-made routines that are part of the
CubExtractor software package (that will be presented in detail in a
companion paper; Cantalupo, in prep.) aiming
at improving flat-fielding and sky subtraction as described in more detail below.
The MUSE pipeline standard recipes (\emph{scibasic} and \emph{scipost}) 
included bias subtraction, initial flat-fielding, wavelength and flux calibration, 
in addition to the geometrical cube reconstruction using the appropriate geometry table 
obtained in our GTO run. 
We did not perform sky-subtraction using the pipeline as we used the sky for each
exposure to improve flat-fielding as described below. 

These initial steps resulted in 36 datacubes, which we registered to the same
frame correcting residual offsets using the positions of sources in the white-light images
obtained by collapsing the cubes in the wavelength direction. 
As commonly observed after the standard pipeline reduction, the white-light images showed
significant flat-fielding residuals and zero-levels fluctuations up to 1\% of the average sky value
across different Integral Field Units (IFUs).
These residuals are both wavelength and flux dependent.
In a companion paper describing the CubExtractor package (Cantalupo, in prep.) 
we discuss the possible origin of these variations and provide more details and test cases
for the procedures described below. 

\subsection{CubeFix: flat-fielding improvement with self-calibration}

 Because our goal is to detect faint and extended emission to levels that are comparable
 to the observed systematic variations, we developed a post-processing routine 
 called CubeFix to improve the flat-fielding by self-calibrating the cube using the observed sky.
 In short, CubeFix calculates a chromatic and multiplicative correction factor that needs to
 be applied to each IFU and to each slice\begin{footnote}{the individual element of an IFU corresponding to a single
 ``slit''.}\end{footnote} within each IFU in order to make the
 measured sky values consistent with each other over the whole Field of View (FoV) of MUSE.
 
 This is accomplished by first dividing the spectral dimension in an automatically obtained
 set of pseudo medium-bands (on sky continuum) and pseudo narrow-bands on sky lines. 
 This is needed both to ensure that there is enough signal to noise in each band for a proper correction
 and to allow the correction to be wavelength and flux dependent. 
 In this step, particular care is taken by the software to completely include all the flux of the 
 (possibly blended) sky lines in the narrow-bands, as line-spread-function variations 
 (discussed below) make the shape and flux density of sky lines vary significantly across the field. 
 Then for each of these bands an image is produced by collapsing the cube along the wavelength dimension.
 A mask (either provided or automatically calculated) is used to exclude continuum sources.
 By knowing the location of the IFU and slices in the MUSE FoV (using the information stored in the
 \emph{pixtable}), CubeFix calculates the averaged sigma clipped values of the sky 
 for each band, IFU and slice and correct these values in order to make them
 as constant as possible across the MUSE FoV. When there are not enough pixels for a slice-to-slice
 correction, e.g. in the presence of a masked source, an average correction is applied using
 the adjacent slices. 
 The slice-by-slice correction is only applied using the medium-bands that include typically
 around 300 wavelength layers each. An additional correction on the IFU level only is then
 performed using the narrow-bands on the sky-lines. This insures that the sky signal
 always dominates with respect to pure line emission sources
  and therefore that these sources do not cause overcorrections.
  We have verified and tested this by injecting fake extended line emission sources 
  with a size of $20\times20$ arcsec$^{2}$ in a single layer at the expected wavelengths of 
  $\ion{He}{ii}$ and $\ion{C}{iv}$ emission of the Slug nebula, both located
  far away in wavelength from skylines, with a SB of $10^{-18}$ erg s$^{-1}$ cm$^{-2}$ arcsec$^{-2}$.
  After sky subtraction, the flux of these sources is recovered within a few percent of the original value.
  We note, however, that caution must be taken when selecting the
  width of the skyline narrow-bands if very bright and extended emission lines are expected 
  to be close to the skylines.
 
  In order to reduce possible overcorrection effects due to continuum sources,
   we performed the CubeFix step iteratively, repeating the procedure after a first total 
   combined cube is obtained (after sky subtraction with CubeSharp as described below). 
 The higher SNR of this combined cube allows a better masking of sources for each individual cube, 
 significantly improving overcorrection problems around very bright sources. 
 We stress that, by construction, a self-calibration method such as CubeFix can only
 work for fields that are not crowded with sources (e.g., a globular cluster) or
 filled by extended continuum sources such as local galaxies. 
 CubeFix has been successfully applied providing excellent results for both quasar 
 and high-redshift galaxy fields
  \citep[including, e.g.,][]{borisova16,fumagalli16,north17,fumagalli17,farina17,ginolfi18,marino18}.
 
 \subsection{CubeSharp: flux conserving sky-subtraction}
 
 The Line Spread Function (LSF) is known to vary both spatially and spectrally
 across the MUSE FoV and wavelength range (e.g., \citealt{bacon15}). Moreover, 
 because of the limited spectral resolution of MUSE, the LSF is typically 
 under-sampled. Temporal variations due to, e.g. temperature changes
 during afternoon calibrations and night-time observations, 
 result in large slice-by-slice fluctuations of sky line fluxes
 in each layer that cannot be corrected by the MUSE standard pipeline 
 method (see e.g., \citealt{bacon15} for an example).
 To deal with this complex problem, other methods have been developed, 
 e.g. based on Principal Component Analysis (ZAP, \citealt{zap}), to reduce the 
 sky subtraction residuals. However, once applied to a datacube with improved flat-fielding
 obtained with CubeFix, the PCA method tends to reintroduce again significant 
 spatial fluctuations. This is because such a method is not necessarily flux conserving
 and the IFUs with the largest LSF variations do not contribute enough to the 
 variance to be corrected by the algorithm. As a result, the layers at the edge of sky lines
may show large extended residuals that mimic extended and faint emission. 

For this reason, we developed an alternative and fully flux-conserving sky-subtraction
method, called CubeSharp, based on an empirical LSF reconstruction 
using the sky-lines themselves. The method is based on the assumption that the sky lines
should have both the same flux and the same \emph{shape} independent on their position in the MUSE FoV. 
After source masking and continuum-source removal, the sky-lines are identified automatically 
and for each of them (or group of them), an average \emph{shape} is calculated
using all unmasked spatial pixels (spaxels). Then, for each spaxel, the flux in each spectral pixel is
moved across neighbours producing flux-conserving LSF variation (both in centroid and width)
until chi-squared differences are minimised with respect to the \emph{average} sky spectrum. 

The procedure is repeated iteratively and it is controlled by several user-definable parameters
that will be described in detail in a separate paper (Cantalupo, in prep.). 
Once these \emph{shifts} have been performed
and the LSFs of the sky lines are similar across the whole MUSE FoV for each layer, sky subtraction
can be performed simply with an average sigma clip for each layer. 
We stress that the method used by CubeSharp could produce artificial line shifts 
by a few pixels (i.e., a few $\mathrm{\AA}$) for line emission close to sky-lines
if not properly masked, however their flux should not be affected (see also the tests of CubeSharp
in \citealt{fumagalli17_Ha}). 
Because the expected line emissions from the Slug nebula do not overlap with sky lines,
this is not a concern for the analysis presented here.

\begin{figure*}
\includegraphics[width=2\columnwidth]{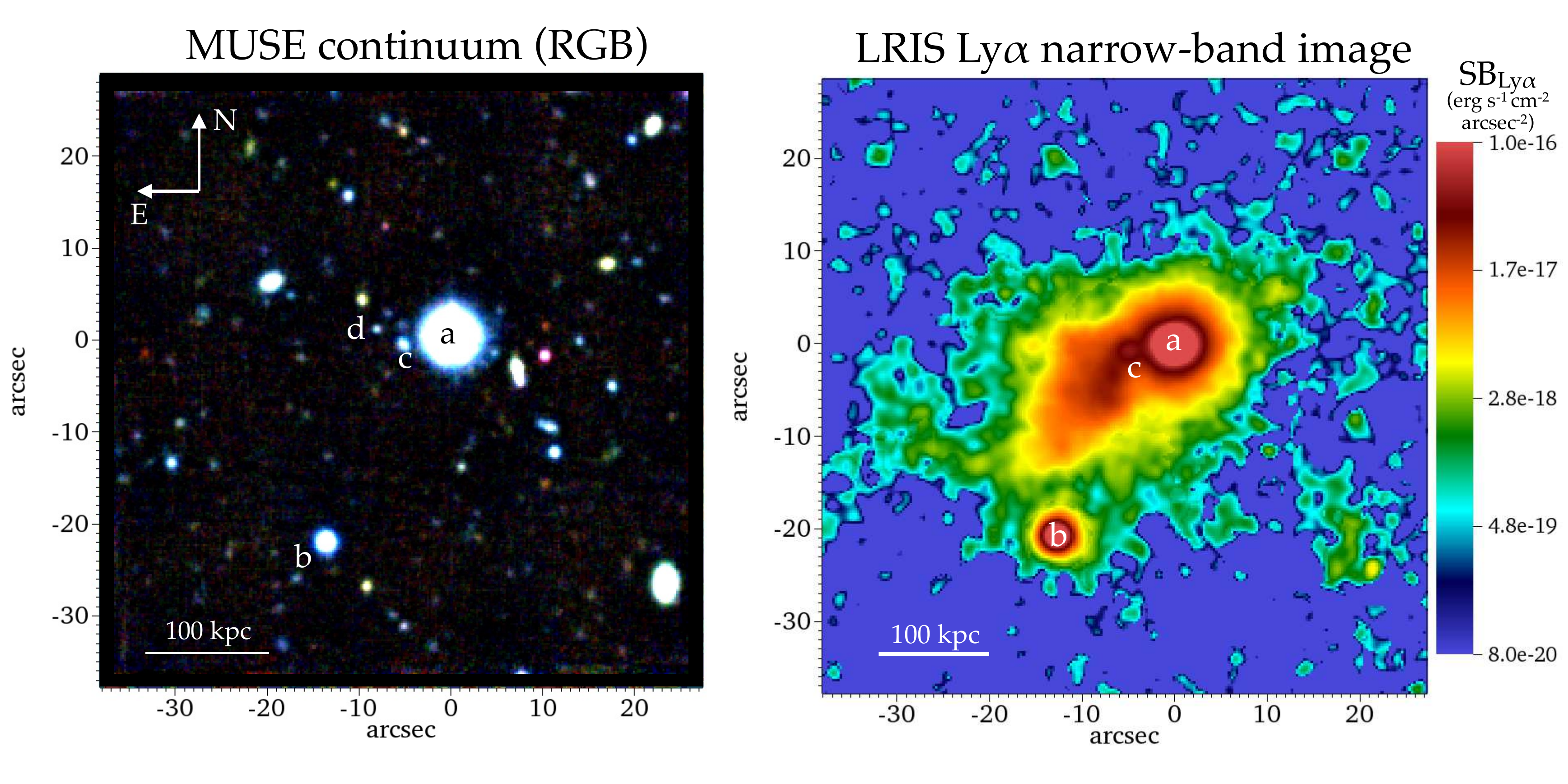}
\caption{
\emph{Left panel}: reconstructed three-colour image of the final 9 hour exposure MUSE datacube
centered on the Slug nebula. This image has been obtained 
by collapsing the cube in the wavelength dimension in three different pseudo-broad-bands:
i) ``blue'' ($4875\mathrm{\AA}-6125\mathrm{\AA}$),
ii) ``green'' ($6125\mathrm{\AA}-7375\mathrm{\AA}$),
iii) ``red'' ($7375\mathrm{\AA}-8625\mathrm{\AA}$).
 {
The bright quasar UM287 ($g\simeq17.5$ AB) and its much fainter quasar companion ($g\simeq23$ AB) 
}
are labelled, respectively as ``a'' and ``b''
in the figure. Two of the brightest continuum sources embedded in the nebula are labelled as
``c'' and ``d'' (this is the same nomenclature as used in \citealt{leibler18}). 
Note that the color-scale is highly saturated in order to better visualise the faintest emission.
\emph{Right panel}: Ly$\alpha$ narrow-band image (reproduced from \citealt{sc14})
of the same field of view as presented in the left panel. In addition to the quasar ``a'' and ``b'',
source ``c'' also shows the presence of enhanced Ly$\alpha$ emission with respect to the
extended nebula.  See text for a detailed discussion of the properties of these sources.
}
\label{FigRGB}
\end{figure*}

\subsection{Cube combination}  

After applying CubeFix and CubeSharp to each individual exposure, a first combined cube is obtained 
using an average sigma clipping method (CubeCombine tool). This first cube is then used
to mask and remove continuum sources in the second iteration of CubeFix and CubeSharp. 
After this iteration, the final cube is obtained with the same method as above. 
The final, combined cube has a 1$\sigma$ noise level of about $10^{-19}$ erg s$^{-1}$ cm$^{-2}$
arcsec$^{-2}$ per layer
in an aperture of 1 arcsec$^{2}$ at 5300$\mathrm{\AA}$, 
around the expected wavelength of the Slug $\ion{He}{ii}$ emission. 

In the left panel of Fig.\ref{FigRGB}, we show an RGB reconstructed image obtained 
by collapsing the cube in the wavelength dimension in three different pseudo-broad-bands:
i) ``blue'' ($4875\mathrm{\AA}-6125\mathrm{\AA}$),
ii) ''green'' ($6125\mathrm{\AA}-7375\mathrm{\AA}$),
iii) ''red'' ($7375\mathrm{\AA}-8625\mathrm{\AA}$),
and by combining them into a single image. 
The 1$\sigma$ continuum noise levels in an aperture of 1 arcsec$^{2}$ are 
$0.33$, $0.27$, $0.33$ in units of 
 { $10^{-20}$ erg s$^{-1}$ cm$^{-2}$ arcsec$^{-2}$ $\mathrm{\AA}$ } 
for the ``blue'', ``green'' and ``red'' pseudo-broad-bands, respectively 
(these noise levels corresponds to AB magnitudes of about 30.1, 29.9, and 29.3 for the same bands).  
 {
The bright quasar UM287 ($g\simeq17.5$ AB) and its much fainter quasar companion ($g\simeq23$ AB) 
}
are labelled, respectively as ``a'' and ``b'' in the figure. Two of the brightest continuum sources 
embedded in the nebula are labelled as ``c'' and ``d'' (this is the same nomenclature as used
in \citet{leibler18}. Source ``c'' shows also
associated compact Ly$\alpha$ emission (see the right panel of Fig.\ref{FigRGB}). 
In section \ref{sec_ol}, we will discuss the properties of these sources in detail.

\section{Analysis and Results}\label{sec_res}

Before the extraction analysis of possible extended emission line associated with the
Slug, we subtracted both the main quasar (UM287) PSF and continuum from 
all the remaining sources, as described below. 

\subsection{QSO PSF subtraction}

Quasar PSF subtraction is necessary in order to disentangle extended line emission from the
line emission associated with the quasar broad line region. Although UM287 does not show
$\ion{He}{ii}\lambda1640$ in emission, we choose to perform the quasar PSF subtraction on the whole available
wavelength range to help the possible detection of other extended emission lines such as, $\ion{C}{iv}$ or $\ion{C}{iii}$
that are also present in the quasar spectrum. 
As for other MUSE quasar observations (both GTO and for several other GO programs) 
PSF subtraction was obtained with CubePSFSub (also part of the CubExtractor package)
based on an empirical PSF reconstruction method (see also \citealt{husemann13} for a 
similar algorithm).
In particular, CubePSFSub uses pseudo-broad-band images of the quasar and its surroundings
and rescales them at each layer under the assumption that the central pixel(s) in the PSF
are dominated by the quasar broad line region. Then the reconstructed PSF is subtracted
from each layer. For our analysis, we used a spectral width of 150 layers for the pseudo-broad-bands
images. We found that this value provided a good compromise between capturing wavelength PSF
variations and obtaining a good signal-to-noise ratio for each reconstructed PSF. 
We limited the PSF corrected area to a maximum distance of about 5'' from the quasar
to avoid nearby continuum sources (see Fig.\ref{FigRGB}) 
from compromising the reliability of our empirically reconstructed PSF.
 {To avoid that the empirically reconstructed PSF could be affected by extended nebular
emission (producing over subtraction), we do not include the range of layers where extended emission is expected.
Because, we do not know a priori in which layers extended emission may be present, 
we run CubePSFSub iteratively, increasing the numbers of masked layers until we obtain a
PSF-subtracted spectrum that has no negative values at the edge of any detectable, 
residual emission line}. 
In particular, in the case of $\ion{He}{ii}\lambda1640$ the masked layers range 
between the number 504 and 516 in the datacube (corresponding to the wavelength range 
5380$\mathrm{\AA}-5395\mathrm{\AA}$).
 {We note that the continuum levels of UM287 at the expected wavelength range of 
$\ion{He}{ii}\lambda1640$ are in any case negligible with respect to the sky-background noise 
at any distance similar or larger than the position of ``source c" (see, e.g., the left panel of Fig.\ref{FigRGB}). 
Therefore, including the PSF subtraction procedure at the $\ion{He}{ii}\lambda1640$ wavelength 
before continuum subtraction (as described below) does not have any noticeable effects on 
the results presented in this work}. 

\subsection{Continuum subtraction}

Continuum subtraction was then performed with CubeBKGSub (also part of the CubExtractor package)
by means of median filtering, spaxel by spaxel, along the spectral dimension using a bin size of 40 pixel
and by further smoothing the result across four neighbouring bins. Also in this case, spectral regions with
signs of extended line emission were masked before performing the median filtering.
In particular, in the case of $\ion{He}{ii}\lambda1640$ we masked every layer between the number 504 and 516 in the datacube
(corresponding to the wavelength range 5380$\mathrm{\AA}-5395\mathrm{\AA}$) as performed during PSF subtraction. 

Finally, we divided the cube into subcubes with a spectral width of about 63$\mathrm{\AA}$ (50 layers) 
around the expected wavelengths of the $\ion{He}{ii}$, $\ion{C}{iv}$, and $\ion{C}{iii}$ emission, i.e., around 5380$\mathrm{\AA}$, 
5080$\mathrm{\AA}$, and 6250$\mathrm{\AA}$, respectively.

\subsection{Three-dimensional signal extraction with CubExtractor}

In order to take full advantage of the sensitivity and capabilities of an integral-field-spectrograph
such as MUSE, three-dimensional analysis and extraction of the signal is essential. 
Intrinsically narrow lines such as the non-resonant $\ion{He}{ii}\lambda1640$ can be detected to very low levels
by integrating over a small number of layers. On the other hand, large velocity shifts due to kinematics,
Hubble flow or radiative transfer effects (in the case of resonant lines) 
could shift narrow emission lines across many spectral layers in different spatial locations.
A single (or a series) of pseudo-narrow bands would therefore either be non-efficient in 
producing the highest possible signal-to-noise ratio from the datacube or missing part of the signal. 

In order to overcome these limitations, we have developed a new three-dimensional extraction
and analysis tool called CubExtractor (CubEx in short) that will be presented in detail in a separate
paper (Cantalupo, in prep.). In short, CubEx performs extraction, detection and (simple) photometry 
of sources with arbitrary spatial and spectral shapes directly within datacubes 
using an efficient connected labeling component algorithm with union finding  
based on classical binary image analysis, similar to the one used by SExtractor \citep{sextractor}, 
but extended to 3D (see e.g., Shapiro \& Stockman, Computer Vision, Mar 2000). 
Datacubes can be filtered (smoothed) with three-dimensional gaussian filters before extraction. 
Then datacube elements, called ''voxels'', are selected if their (smoothed) flux is above a user-selected
signal-to-noise threshold with respect to the associated variance datacube. 
Finally, selected voxels are grouped together within objects that are discarded if their number
of voxels is below a user-defined threshold. 
CubEx produces both catalogues of objects (including all astrometric, photometric and spectroscopic 
information) and datacubes in FITS format, including: 
i) ``segmentation cubes'' that can be used to perform further analysis
(see below) and, ii) three-dimensional signal-to-noise cubes of the detected objects 
that can be visualised in three-dimensions with several public visualization softwares 
(e.g., VisIt\begin{footnote}{https://wci.llnl.gov/simulation/computer-codes/visit; see also \citet{VisIt}}\end{footnote}).

\subsection{Detection of extended $\ion{He}{ii}$ emission}\label{HeIIDet}

\begin{figure}
\includegraphics[width=1.05\columnwidth]{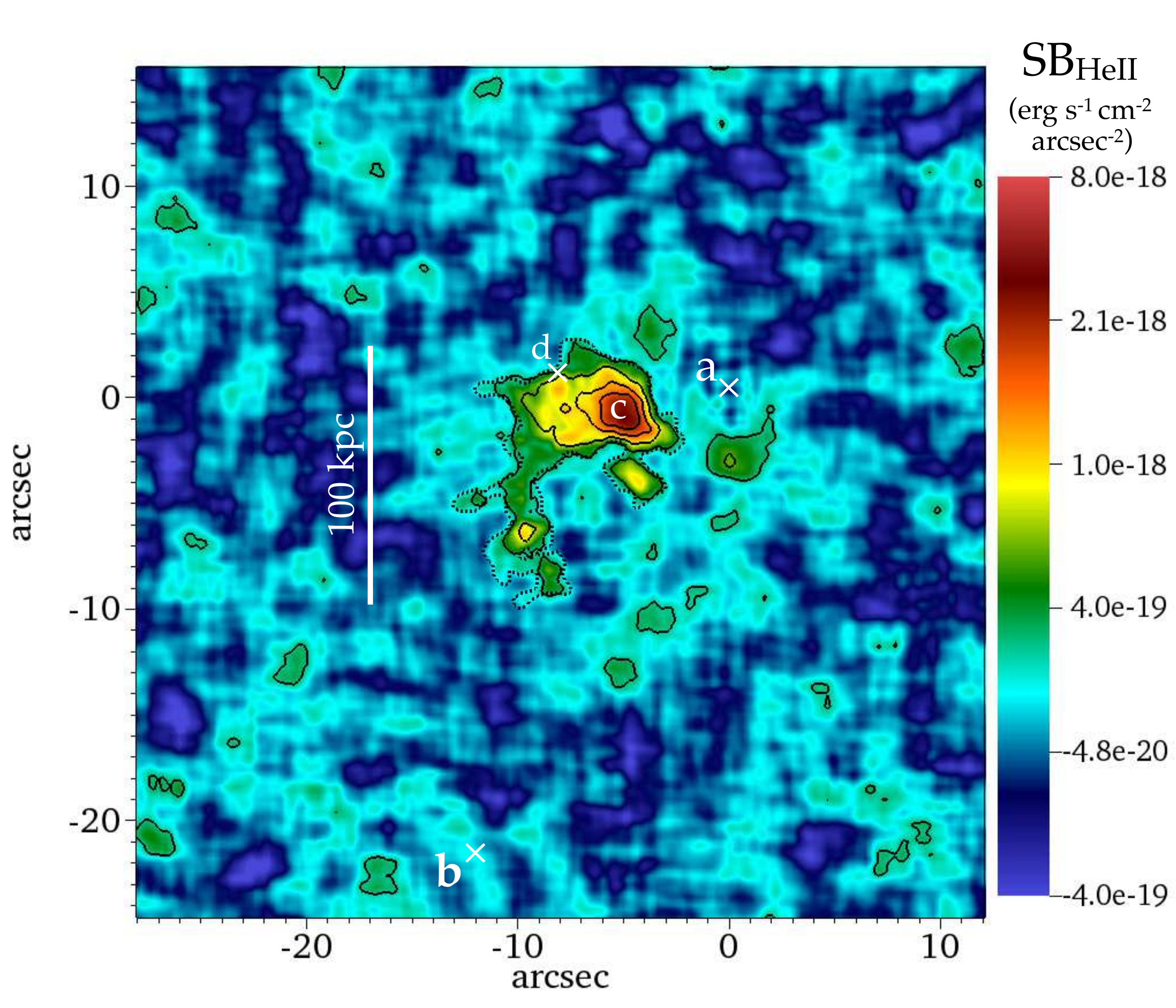}
\caption{
``Optimally extracted'' image of the detected $\ion{He}{ii}$ emission from the Slug nebula.
This image has been obtained by integrating along the
spectral direction the SB of all the voxels associated with this source in the
CubEx ``segmentation cube'' (see text for details). 
These voxels are contained within the overlaid dotted contour. Outside of these
contours (where no voxels are associated with the detected emission)
we show for comparison the SB of the voxels in a single layer 
close to the central wavelength of the detected emission. 
Before spectral integration, a spatial smoothing with size of 0.8" has been applied
to improve visualisation. Solid contours indicate SNR levels of 2, 4, 6, and 8.
The positions of quasars ``a'', ``b'', and sources ``c'' 
and ``d"
are indicated in the figure.
 The brightest $\ion{He}{ii}$ emission - approaching a SB close to 
10$^{-17}$ erg s$^{-1}$ cm$^{-2}$ arcsec$^{-2}$ - is located in correspondence of the compact source ``c''.
The region above a SB of about 10$^{-18}$ erg s$^{-1}$ cm$^{-2}$ arcsec$^{-2}$ 
or SNR$>$4 
(third solid contour line around source ``c")
 covers a projected area of about 6''$\times$3.5" (or about 50$\times$30 physical kpc).
 We refer to this region as ``region c'' in the text
 {(see also Fig.\ref{Fig_LyaSB_HeII})}
}
\label{FigHeII_OE}
\end{figure}

We run CubEx on the subcube centered on the expected $\ion{He}{ii}$ emission with the following parameters: 
i) automatic rescaling of the pipeline propagated variance\begin{footnote}{It is known that the variance in the
MUSE datacubes obtained from the pipeline tends to be underestimated by about a factor of two
due to, e.g. resampling effects (see e.g., \citealt{bacon17}). We rescale the variance layer-by-layer with CubEx 
using the following procedure: i) we compute the variance of the measured flux between spaxels in each
layer (``empirical variance''), ii) we rescale the average variance in each layer in order to match the ``empirical 
variance'', iii) we smooth the rescaling factors across neighbouring layers to avoid sharp transitions due to,
e.g. the effect of sky line noise.}\end{footnote},
ii) smoothing in the spatial and spectral dimension with a gaussian kernel of radius 
of 0.4'' and 1.25$\mathrm{\AA}$ respectively,
iii) a set of signal-to-noise (SNR) threshold ranging from 2 to 2.5,
iv) a set of minimum number of connected voxels ranging from 500 to 5000.
In all cases, we detected at least one extended source with more than 5000 connected voxels above 
a SNR threshold of 2.5. This source - that we call ''region c'' - 
is located within part of the area covered by the Slug Ly$\alpha$ 
emission and, in particular, overlaps with sources ``c'' and ``d'' 
(see Fig.\ref{FigHeII_OE}). However, it does not cover the area occupied
by the brightest Ly$\alpha$ emission - that we call ``bright tail'' - that extends  south of source ``c'' 
by about 8'' (see the right panel in Fig.\ref{FigRGB}) at any explored SNR levels. 
This result does not change if we modify our spatial smoothing radius or
do not perform smoothing in the spectral direction.
The other detected source is the  {spatially compact but spectrally broader $\ion{He}{ii}$ emission associated with the 
broad-line-region of faint quasar ``b'' (not shown in Fig.\ref{FigHeII_OE} ) while there is no clear
detection within 2" of quasar ``a".}
Moreover, we have
no information on the presence of nebular 
Ly$\alpha$ emission in this region because of the difficulties
of removing the quasar PSF from the LRIS narrow-band imaging
 {
\begin{footnote}{this is due to the fact that the LRIS narrow-band 
and continuum filter changes significantly across the FOV
and because of the brightness of UM287 in both Ly$\alpha$
and continuum. Unfortunately, it was not possible for us to find a
nearby, unsaturated and isolated star with similar brightness of UM287 
and close enough to the position of the quasar 
to obtain a good empirical estimation of the PSF for the correction
using a simple rescaling factor.
}\end{footnote}
}
. For these reasons, 
we cannot reliably constrain the $\ion{He}{ii}$/Ly$\alpha$ ratio within a few arcsec from quasar
``a" and we will not consider this region in our discussion. 
 {
In section \ref{secLineratio} we estimate an upper limit to the possible contribution
of the quasar Ly$\alpha$ emission PSF to the regions of interest in this work.
}
Other, much smaller objects that appeared at low SNR thresholds
are likely spurious given their morphology.  
To be conservative, we used in the rest of the $\ion{He}{ii}$ analysis of the ``region c'' the segmentation 
cube obtained by CubEx with a SNR threshold of 2.5.

In Fig. \ref{FigHeII_OE}, we show the ``optimally extracted'' image
of the detected $\ion{He}{ii}$ emission obtained by integrating along the
spectral direction the SB of all the voxels associated with this source in the
CubEx segmentation cube. 
These voxels are contained within the overlaid dotted contour. Outside of these
contours (where no voxels are associated with the detected emission)
we show for comparison the SB of the voxels in a single layer 
close to the central wavelength of the detected emission. 
Before spectral integration, a spatial smoothing with size of 0.8" has been applied
to improve the visualisation. 
We stress that the purpose of this optimally extracted image (obtained with the tool Cube2Im) 
is to maximise the signal to noise ratio of the detection rather than the flux.
However, by growing the size of the spectral region used for the integration, we have
verified that the measured flux in the optimally extracted image can be considered
a good approximation to the total flux within the measurement errors.
This is likely due to the fact that we are smoothing also in the spectral direction
and that the line is spectrally narrow as discussed in section \ref{seckin}.
We note that the brightest $\ion{He}{ii}$ emission - approaching a SB close to 
10$^{-17}$ erg s$^{-1}$ cm$^{-2}$ arcsec$^{-2}$ - is located in correspondence of the compact source ``c''.
The region above a SB of about 10$^{-18}$ erg s$^{-1}$ cm$^{-2}$ arcsec$^{-2}$ 
(coloured yellow in the figure) extends by about 5'' (or about 40 kpc) in the direction of 
source ``d''. The overall extension of the detected region approaches 12'', i.e., about 100 kpc.
 {Below but still connected with this region there is a ``faint tail" of emission 
detected with SNR between 2.5 and 4. Because the significance of this emission 
is lower, to be conservative we will focus in our discussion on 
the high SNR part of the emission (``region c").}

\begin{figure}
\includegraphics[width=\columnwidth]{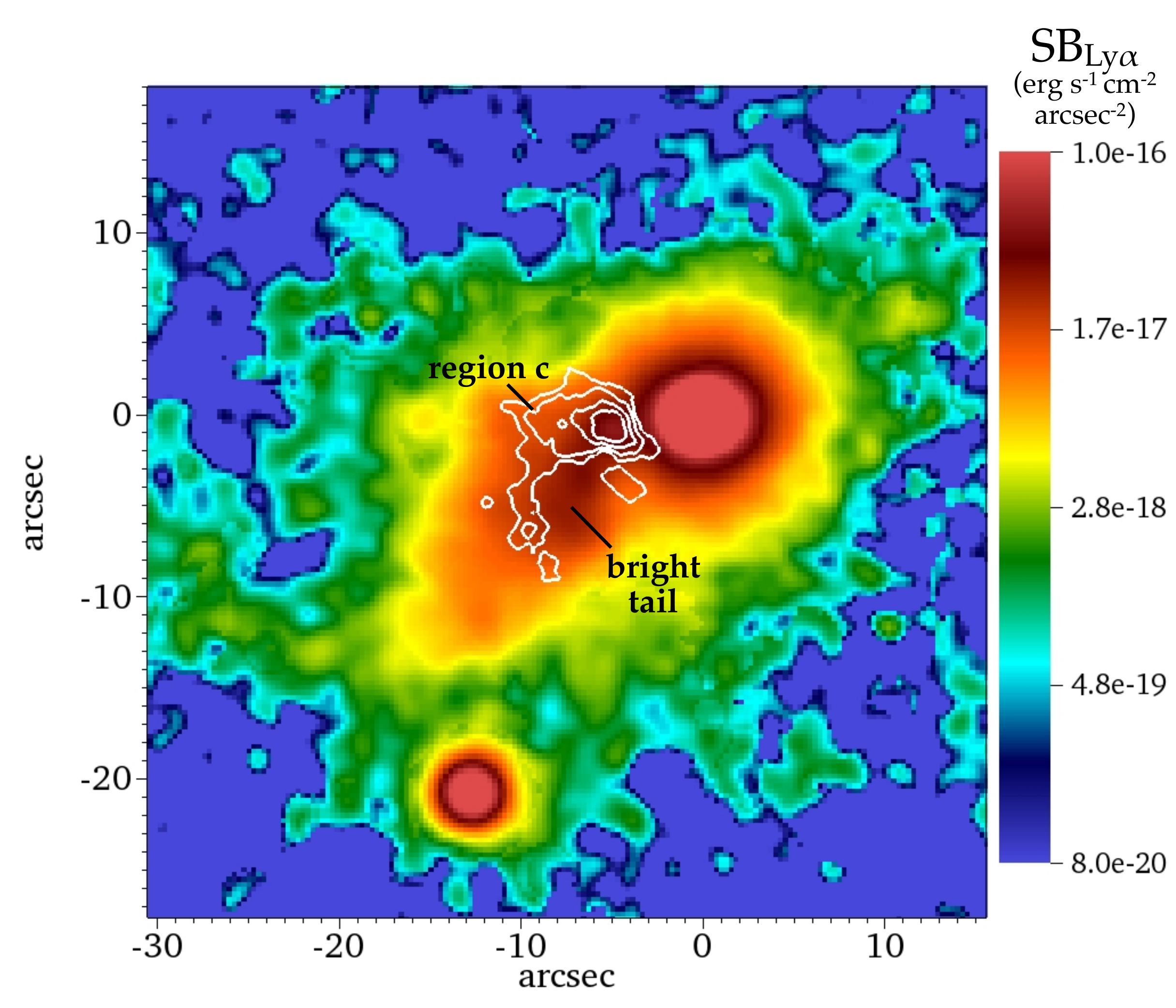}
\caption{
SNR contours of the detected $\ion{He}{ii}$ emission (solid white lines; see Fig.\ref{FigHeII_OE})
overlaid on the Ly$\alpha$ narrow-band image presented in right panel
of Fig.\ref{FigRGB}.
There is little correspondence between the location of the 
brightest Ly$\alpha$ emission (that we call ''bright tail'', roughly centred at position $\Delta x=-7$" and $\Delta y=-5$"
and labeled in the figure; 
we will use the Ly$\alpha$ surface brightness at this position in our calculations 
of line ratios for the ``bright tail"
) 
and the majority of the $\ion{He}{ii}$ emission. The exception is the position occupied 
by the compact source ``c''. Indeed, the $\ion{He}{ii}$ emitting region seems to avoid the ''bright tail''. 
}
\label{Fig_LyaSB_HeII}
\end{figure}

In Fig.\ref{Fig_LyaSB_HeII}, we overlay the SNR contours 
of the detected $\ion{He}{ii}$ emission
on the Ly$\alpha$ image for a more direct comparison. These contours have been
obtained by propagating, for each spaxel, the estimated (and rescaled) variance from the
pipeline (see section \ref{HeIIDet}) taking into account the numbers of layers that
contribute to the ``optimally extracted" image in that spatial position (see also \citealt{borisova16}).
As is clear from Fig. \ref{Fig_LyaSB_HeII},
there is very little correspondence between the location of the brightest Ly$\alpha$ emission
(''bright tail'', 
 {labeled in the figure}
) and the majority of the $\ion{He}{ii}$ emission, with the exception of the 
exact position occupied by the compact source ``c''. 
Indeed, the $\ion{He}{ii}$ region seems to avoid the ''bright tail''. We will 
present in section \ref{secLineratio} the implied line ratios and we will 
explore in detail the implications of this result in the discussion section.

\subsection{Kinematic properties of the $\ion{He}{ii}$ emission}\label{seckin}

\begin{figure}
\includegraphics[width=\columnwidth]{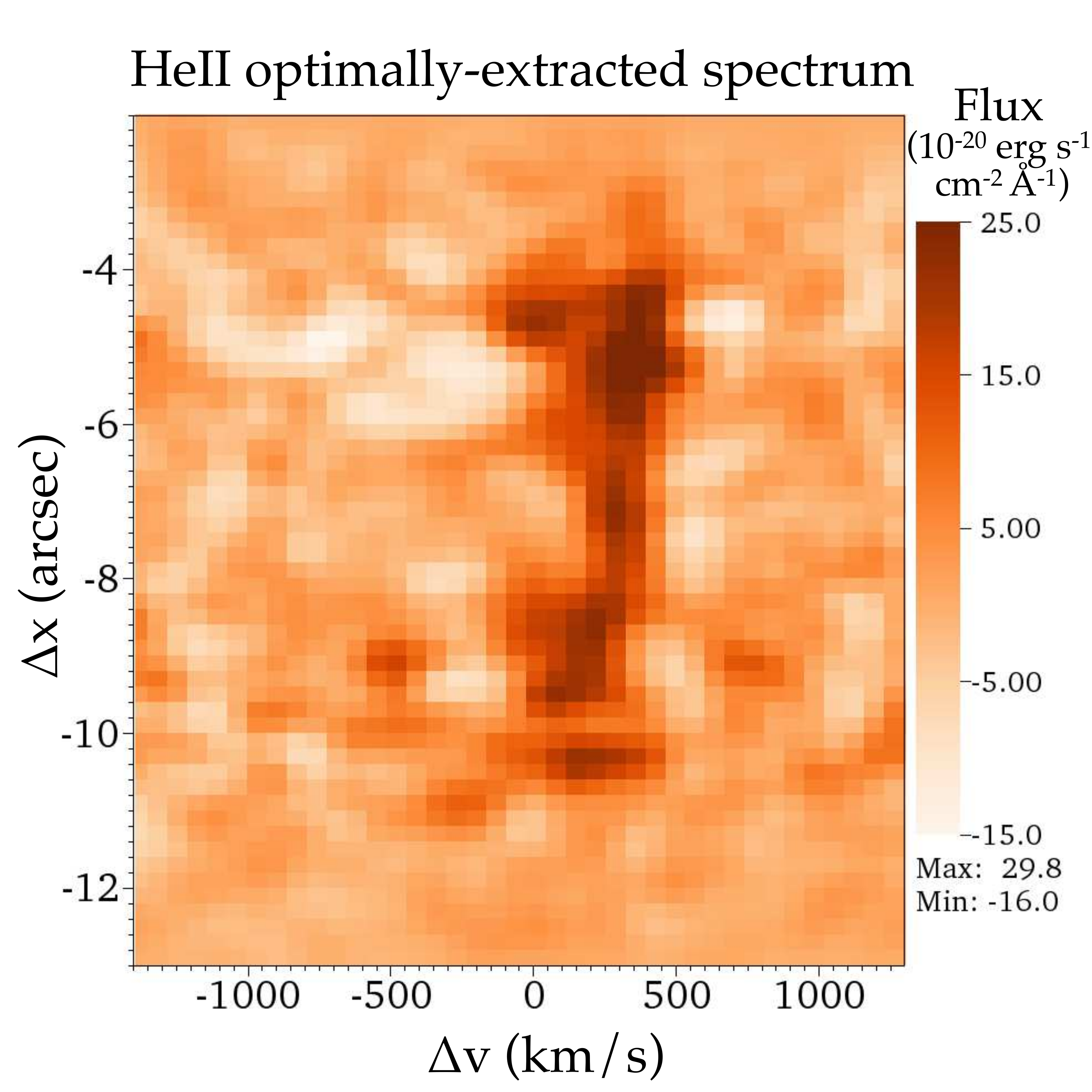}
\caption{
``Optimally extracted''
two-dimensional spectrum of the detected $\ion{He}{ii}$ emission
projected along the y-axis direction of Fig. \ref{FigHeII_OE}.
In particular, this spectrum has been obtained using the 
``segmentation cube'' produced by CubEx (see text for details). 
Zero velocity corresponds to the CO systemic redshift of quasar ``a''  
(i.e., z$=2.283$, DeCarli et al. in prep.) and the y-axis represents the projected 
distance (along the right ascension direction, i.e.
the x-axis in the previous figures) in arcsec from ``a''.
For visualisation purposes, we have smoothed the cube in the spatial direction 
with a gaussian with radius 1 pixel (0.2") before extracting the spectrum.
}
\label{Fig_HeII_OESpc}
\end{figure}

\begin{figure}
\includegraphics[width=\columnwidth]{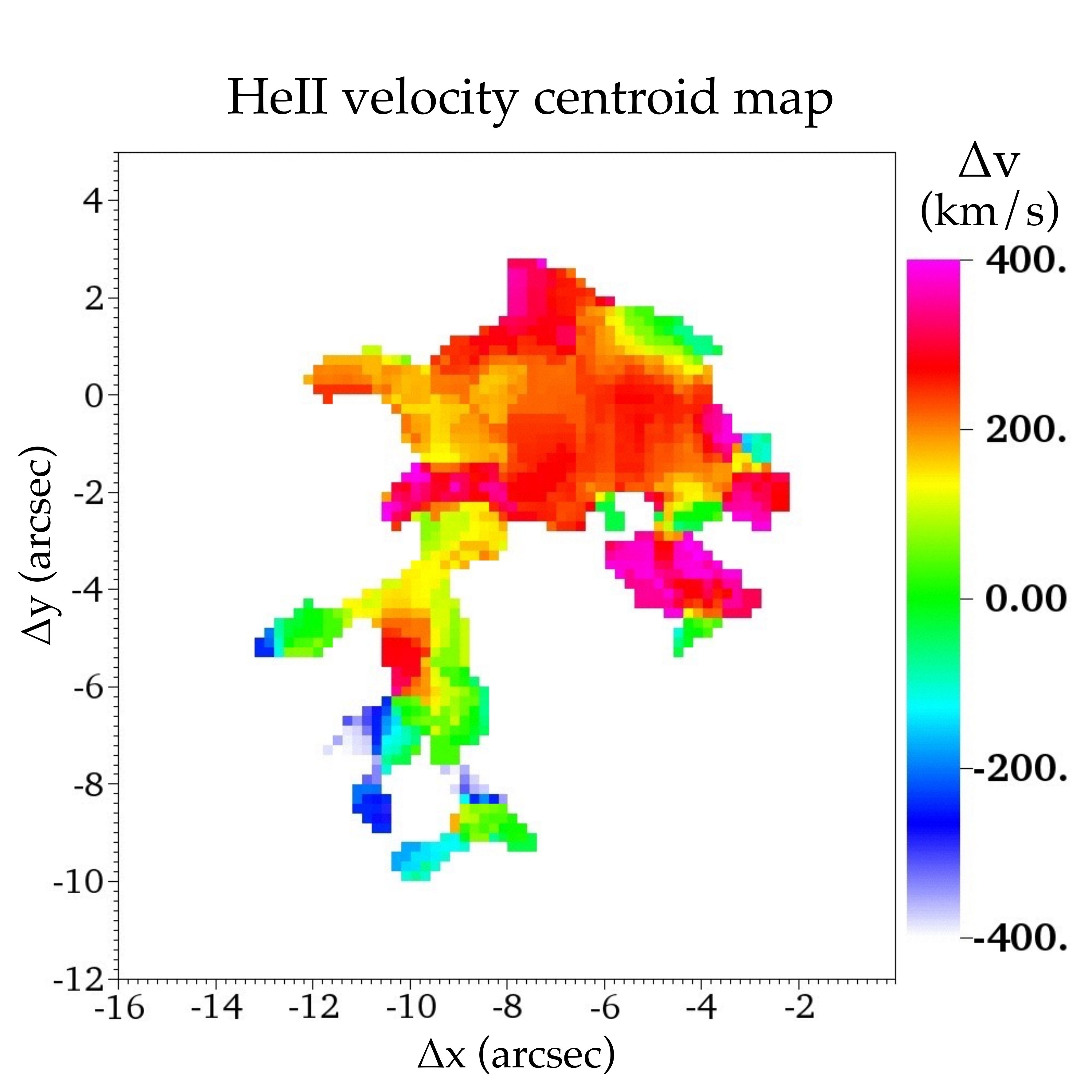}
\caption{
Two-dimensional map of the $\ion{He}{ii}$ emission velocity centroid, 
obtained as the first moment of the flux distribution of the voxels associated with
the source detected by CubEx. The majority of the emission,
located between $\Delta y=-3$'',$\Delta y=3$'' (``region c"),
shows a typical velocity shift between 200 and 300 km/s from the 
systemic redshift of quasar ``a'' and no evidence for ordered 
kinematical patterns such as, rotation, inflows or outflows. 
We note that the velocity shift of about 200km/s in the ''region c''
is remarkably close the the one measured both in Ly$\alpha$ 
and H$\alpha$ emission at the same spatial location (see \citealt{leibler18}).
}
\label{Fig_HeII_vmap}
\end{figure}

In Fig.\ref{Fig_HeII_OESpc}, we show the optimally extracted
two-dimensional spectrum of the detected $\ion{He}{ii}$ emission
projected along the y-axis direction. This spectrum has been
obtained in the following way (automatically produced with the 
tool Cube2Im): i) we first calculated the spatial projection of 
the segmentation cube with the voxels associated with the detected object
(``2d mask''; this region is indicated by the dotted contour in Fig.\ref{FigHeII_OE});
ii) we then used this 2d mask as a pseudo-aperture to calculate
the spectrum integrating along the y-axis direction. 
In practice, this procedure maximises the signal-to-noise 
using a matched aperture shape. We notice that for each individual 
spatial position (vertical axis in the two-dimensional spectrum), 
the same number of voxels contribute to the flux, independent of 
the spectral position. However, the number of 
contributing voxels, and therefore the associated noise, 
may change between different spatial positions (as apparent in Fig.\ref{Fig_HeII_OESpc}).
We used as zero-velocity the systemic redshift of the bright quasar ``a'' 
obtained by CO measurements (i.e., z$=2.283$, DeCarli et al. in prep.)
and the y-axis represents the projected distance (along the right ascension direction, i.e.
the x-axis in the previous figures) in arcsec from ``a''.

The detected emission clearly stands-out along the spectral direction at
high signal-to-noise levels between $\Delta x=-4$" and $\Delta x=-11$"
and it is mostly centered around $\Delta v=300$km/s with
coherent kinematics (at least in the region between $\Delta x=-4$" and $\Delta x=-8$").
Moreover, the emission appears very narrow in the spectral direction,
despite the fact that we are integrating along about 4'' in the y-spatial-direction. 
In particular, the FWHM in the central region ($\Delta x=-7$") is  
only about 200 km/s, without deconvolution with the instrumental LSF,
i.e., the line is barely resolved in our observation. 

In Fig. \ref{Fig_HeII_vmap}, we show the two-dimensional map of the
velocity centroid of the emission, obtained as the first moment of the
flux distribution (using the tool Cube2Im) of the voxels associated with
the detected source. As in Fig.\ref{Fig_HeII_OESpc}, the majority of the emission,
located between $\Delta y=-3$" and $\Delta y=3$" (``region c''),
shows a typical velocity shift between 200 and 300 km/s from the 
systemic redshift of quasar ``a'' with a remarkable coherence across
distant spatial locations (with the exception of few regions associated with low 
signal-to-noise emission). 
At least at the spectral resolution of our 
observations, there is no evidence of ordered kinematical patterns
such as, rotation, inflows or outflows. 
The lower signal-to-noise part of the emission located below 
$\Delta y=-3$" seems to show instead a velocity consistent with the
systemic redshift of quasar ``a'' with large variations probably due to noise.  
 
 We note that the velocity shift of about  {300km/s} in this ``region c''
 is remarkably close the the velocity shift measured both in Ly$\alpha$ 
and H$\alpha$ emission in the same spatial location 
 {(about 250 km/s for Ly$\alpha$ and about 400km/s for H$\alpha$; see Figs. 3  and 4 in \citealt{leibler18}).
We also note that the Ly$\alpha$ emission appears broader 
(with a velocity dispersion of about 250km/s) and more asymmetric than the
$\ion{He}{ii}$ emission, as expected in presence of radiative transfer effects.}

\subsection{$\ion{He}{ii}$ and Ly$\alpha$ line ratios}\label{secLineratio}

\begin{figure}
\includegraphics[width=\columnwidth]{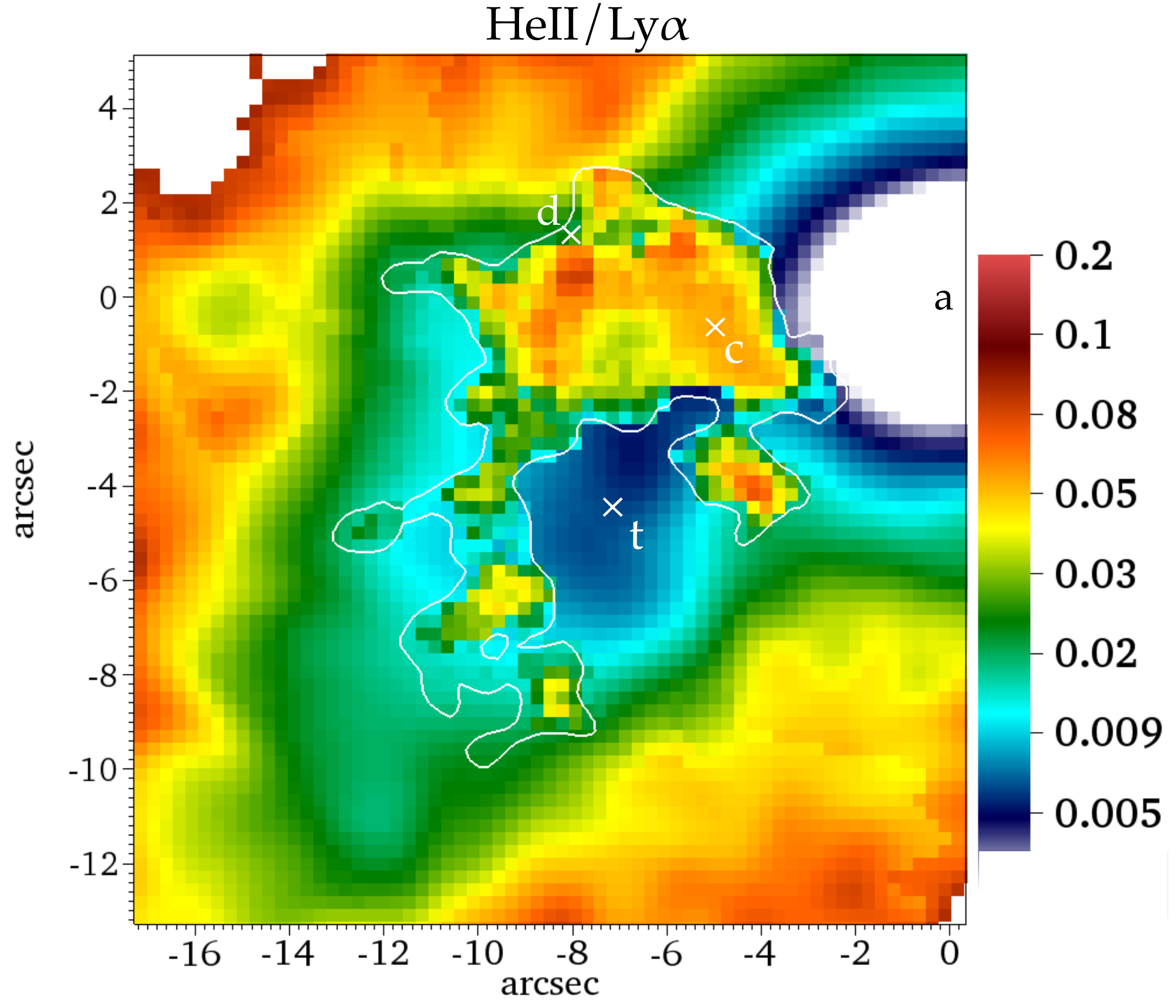}
\caption{
Two-dimensional $\ion{He}{ii}$/Ly$\alpha$ ratio map. The region within the white contours
represents measured values while the rest indicates 
1$\sigma$ upper limits in an aperture of 0.8$\times$0.8 arcsec$^{2}$ and spectral 
width of 3.75$\mathrm{\AA}$. White colour 
indicates regions with no constraints.
 In section \ref{secLineratio} we
describe the detailed procedure used to obtain this map. 
 The two adjacent and Ly$\alpha$-bright regions close to source ``c'' and the ''bright tail'' 
 immediately below show very different line ratios (or limits): the $\ion{He}{ii}$-detected ''region c'' 
 shows $\ion{He}{ii}$/Ly$\alpha$ up to 5\% close to source ``c'' on average 
 (increasing to about 8\% close to source ``d'') while the region immediately below 
that we call ``bright tail'' (i.e., around $\Delta x\simeq -7$" and
$\Delta y\simeq -5$";
 {
indicated by ``t" in the figure
}
) shows 1$\sigma$ upper limits as low as 0.5\%.
 {
As discussed in \ref{secLineratio}, 
we estimated that the quasar ``a" Ly$\alpha$ PSF 
would at maximum increase the $\ion{He}{ii}$/Ly$\alpha$ ratio by 25\% 
close to source ``c" and by less than 10\% in the ``bright tail" region. 
}
In section \ref{Disc} we discuss the possible origin and implication
of both the large gradients in the line ratio and the low measured values
and upper limits. 
}
\label{FigLR}
\end{figure}

In Fig. \ref{FigLR}, we present the two-dimensional map of the measured
(or 1$\sigma$ upper limit in an aperture of 0.8$\times$0.8 arcsec$^{2}$
and spectral width of 3.75$\mathrm{\AA}$) line ratio between 
$\ion{He}{ii}$ and Ly$\alpha$ emission combining our MUSE observations with our previous 
Ly$\alpha$ narrow-band image \citep{sc14}.
The measured values are enclosed within the white contour while the rest of the 
image represents 1$\sigma$ upper limit because of the lack of $\ion{He}{ii}$ detection
in these regions. 
 {
We note that the values and limits within a few arcsec from quasar ``a" could be
artificially lowered by the effects of the quasar Ly$\alpha$ PSF
(that has not been removed in this image for the reasons 
mentioned in section \ref{HeIIDet}). However, as discussed below,
we estimated that quasar PSF effects would at maximum increase the
$\ion{He}{ii}$/Ly$\alpha$ ratio by 25\% close to source ``c" and 
by less than 10\% in the ``bright tail" region. 
} 
%
We have obtained this two-dimensional line ratio map, 
using the following procedure:
i) we smoothed the cube in the spatial directions
with a boxcar with size 0.8" (4 spaxels), i.e.
the FWHM of the measured PSF;
ii) we obtained an optimally extracted image 
from the smoothed cube as described in section \ref{HeIIDet} 
(as discussed in the same section, this image represents 
the total $\ion{He}{ii}$ flux to within a good approximation);
iii) we measured the average noise properties 
in the smoothed cube integrating within the three wavelength layers
closer to the $\ion{He}{ii}$ emission, obtaining a 1$\sigma$ value of
 $1.69\times10^{-19}$ erg s$^{-1}$ cm$^{-2}$ arcsec$^{-2}$ per smoothed pixel
 (equivalent to an aperture of 0.8"$\times$0.8") and 
 spectral width of 3.75$\mathrm{\AA}$;
 iv) we replaced each spaxel  without detected $\ion{He}{ii}$ emission
 in the optimally extracted image with the 1$\sigma$ noise
 value as calculated above;
v) we resampled the spatial scale of this image to match
the spatial resolution of the LRIS Ly$\alpha$ NB image
(i.e., 0.27" compared to the 0.2" of MUSE);
vi) we extracted the Ly$\alpha$ emission from the LRIS image
using CubEx and replaced pixels without detected emission
with zeros;
vii) we cut the LRIS image to match the astrometric properties
of the MUSE optimally extracted image (using quasar ``a'' and ``b'' as
the astrometric reference);
viii) finally, we divided the two images by each other to obtain
the measured $\ion{He}{ii}$ to Ly$\alpha$ line ratios (within the $\ion{He}{ii}$
detected region) or the line ratio 1$\sigma$ upper limit (in the region where
$\ion{He}{ii}$ was not detected and Ly$\alpha$ is present).

The image presented in Fig. \ref{FigLR} quantifies the large difference
in terms of line ratios between the two adjacent and Ly$\alpha$-bright
regions close to source ``c'' and the ''bright tail'' immediately below.
In particular, the $\ion{He}{ii}$-detected ''region c'' shows $\ion{He}{ii}$/Ly$\alpha$ up
to 5\% close to source ``c'' on average (increasing to about 8\% 
 {
1" south
}
of source ``d'')
while the region immediately below (i.e., around $\Delta x\simeq -7$" and
$\Delta y\simeq -5$", indicated by a ``t") shows 1$\sigma$ upper limits as low as 0.5\%.

 {
We note that these values can only be marginally affected by the 
lack of quasar Ly$\alpha$ PSF removal in our LRIS narrow-band image.
In particular, we have estimated the maximum quasar Ly$\alpha$ PSF 
contribution by assuming that all the Ly$\alpha$ emission on the opposite side of the
quasar position with respect to source ``c"  and the ``bright tail" 
(i.e., emission at position $\Delta x>0$ in right-hand panel of Fig.\ref{FigRGB})
is due to PSF effects. In this extreme hypothesis, we obtain that only about
25\% of the Ly$\alpha$ emission around the location of source ``c"
and less than 10\% of the Ly$\alpha$ emission in the bright tail could be 
affected by the quasar Ly$\alpha$ PSF. This effects would of course 
translate in an increased $\ion{He}{ii}$/Ly$\alpha$ ratio of about 25\% around 
source ``c" and less than a 10\% for the ``bright tail". 
We include these effects in the error bars associated with the
measurements in these regions in the rest of this work.
}

Could the large gradient in line ratios be due to different techniques
used to map $\ion{He}{ii}$ emission (i.e., integral-field spectroscopy) and
Ly$\alpha$ (i.e., narrow band, with its limited transmission window)?
The NB filter used on LRIS is centered on the Ly$\alpha$ wavelength
corresponding to z=2.279 \citep{sc14}, i.e. at a velocity separation
of about $-350$ km/s from the quasar systemic redshift measured
with the CO line that we are using as the reference throughout this paper. 
Therefore the Ly$\alpha$ emission associated
with the ``bright tail'' is located exactly at the peak of the NB transmission 
window (see also \citealt{leibler18}).
 The FWHM of the filter corresponds to about 
3000 km/s, therefore the filter transmission would be about half
the peak value at a shift of about 1150 km/s with 
respect to the quasar ``a'' systemic redshift. 
Both the Ly$\alpha$ and the $\ion{He}{ii}$ emission detected 
from the ``region c'' extend up to a maximum velocity
shift of about 500 km/s (see Fig.\ref{Fig_HeII_vmap}  and \citealt{leibler18}). 
This is well within the high transmission region
of the NB filter and therefore the Ly$\alpha$ SB of 
``region c'' used in this paper could be underestimated by
a factor less than two. This is much smaller than the factor
of at least 10 difference in the observed line ratios.
Therefore we conclude that the different observational techniques
should not strongly affect our results.
We will discuss the implications of the line ratios in terms of physical
properties of the emitting gas in section \ref{Disc}.

\subsection{Other emission lines}\label{sec_ol}

Using a similar procedure to the one applied to detect and extract
extended $\ion{He}{ii}$ emission, we also searched for the presence of 
extended $\ion{C}{iii}$ and $\ion{C}{iv}$ emission (both doublets). The only location within
the Slug where $\ion{C}{iii}$ and $\ion{C}{iv}$ are detected at significant levels 
is in correspondence of the exact position of the compact source ``c''.

\begin{figure}
\includegraphics[width=\columnwidth]{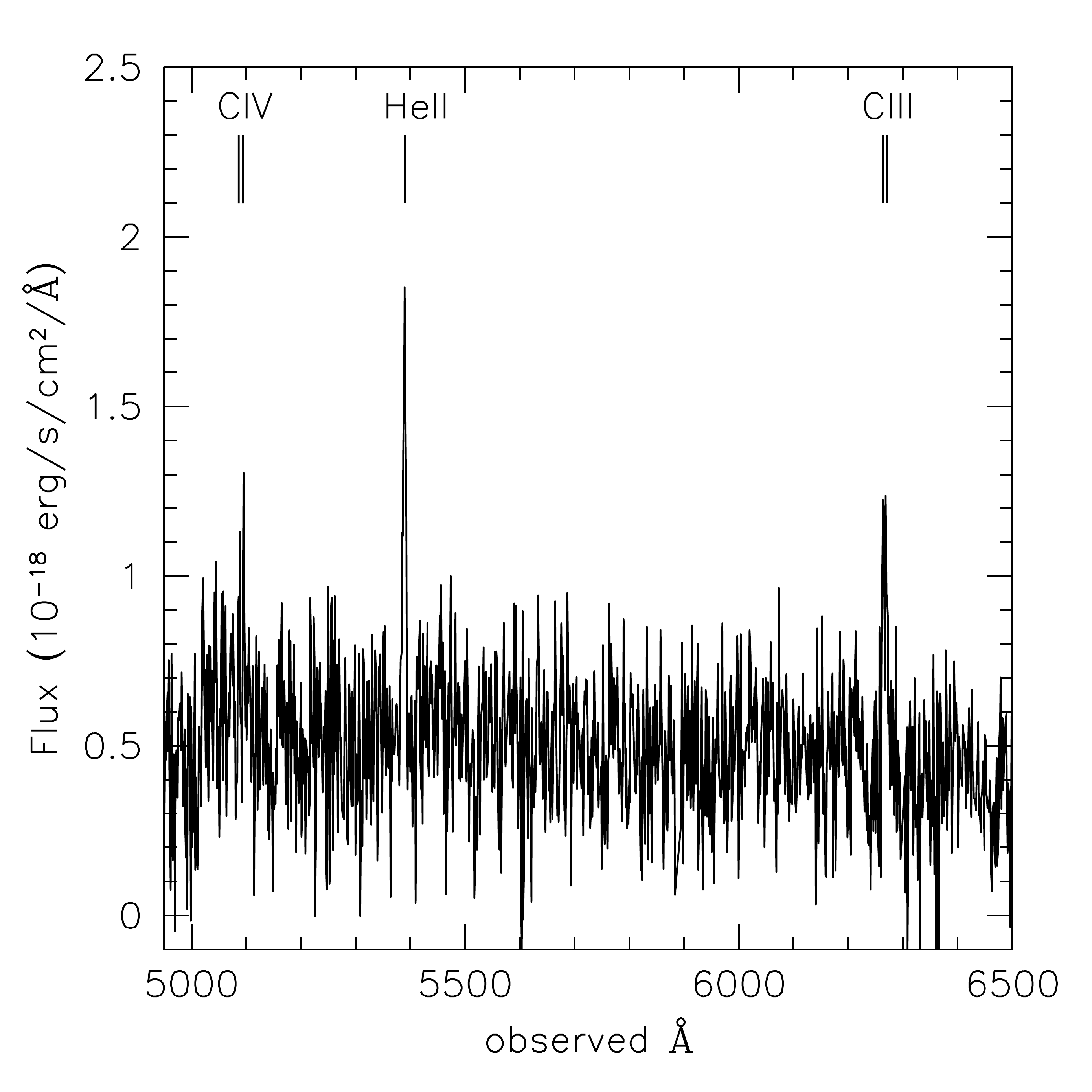}
\caption{
One-dimensional spectrum of the compact ``source c'' obtained within a circular aperture of diameter 
1.6" (about twice the seeing FWHM) before continuum subtraction and after quasar PSF subtraction. 
The expected positions of $\ion{C}{iv}$ and $\ion{C}{iii}$ given the redshift obtained by the $\ion{He}{ii}$ emission
are labeled in the figure. Both the $\ion{C}{iv}$ and $\ion{C}{iii}$ doublets are 
detected above an integrated signal to noise ratio of 3. 
}
\label{Fig_CSpc}
\end{figure}

In Fig.\ref{Fig_CSpc}, we show the one-dimensional spectrum
obtained by integrating within a circular aperture of diameter 
1.6'' (about twice the seeing FWHM) centered on source ``c''
before continuum subtraction and after quasar PSF subtraction. 
The expected positions of $\ion{C}{iv}$ and $\ion{C}{iii}$ given the redshift obtained by the $\ion{He}{ii}$ emission
are labeled in the figure. Both $\ion{C}{iv}$ and $\ion{C}{iii}$ doublets are 
detected above an integrated signal to noise ratio of 3. 
Moreover, their redshifts are both exactly centered, within the measurement errors, 
on the systemic redshift inferred by the $\ion{He}{ii}$ line.
As for $\ion{He}{ii}$, both $\ion{C}{iii}$ and $\ion{C}{iv}$ are very narrow and marginally resolved 
spectroscopically. However, the detected signal to noise is too low in this
case for a kinematic analysis. After continuum subtraction, both $\ion{C}{iii}$ and $\ion{C}{iv}$
have about half of the flux of the $\ion{He}{ii}$ line within the same photometric aperture. 

The continuum has an observed flux density of about 
  { 
5$\times10^{-19}$ erg s$^{-1}$
cm$^{-2}$ $\mathrm{\AA}$ at 5000$\mathrm{\AA}$ (observed) 
and a UV-slope of about $\beta=-2.2$ 
(estimated from the spectrum between the rest-frame region 
1670$\mathrm{\AA}$ to 2280$\mathrm{\AA}$) 
if the spectrum is approximated with a power-law 
defined as $f_{\lambda}\propto \lambda^{\beta}$.
This value of $\beta$ 
}
 would correspond to 
a extremely modest dust attenuation of $E(B-V)\sim0-0.04$
(following \citealt{bouwens14}). For a starburst with an
age between 10 and 250 Myr, the observed flux and $E(B-V)$
would imply a modest star formation rate ranging between
2 and 6 solar masses per year (e.g., \citealt{otifloranes10}).

In addition to the location of source ``c'' we have found some
very tentative evidence (between 1 and 2 $\sigma$ confidence levels) 
for the presence of extended $\ion{C}{iii}$ at the 
 {
spatial
}
location of the ``bright tail'' 
and for the presence of extended $\ion{C}{iv}$ in the ``region c'' after large 
spatial smoothing
 {
 ($>5"$ in size) 
 }
 in a small range of wavelength layers
  {
around the expected position.
} 
Because of the large uncertainty of these possible 
detections, we leave further analysis to future work.
In particular, either deeper data or more specific
tools for the extraction of extended line emission at
very low SNR would be needed.

\section{Discussion}\label{Disc}

We now focus our attention on the following questions: 
i)
what is the origin of the large variations in both the Slug $\ion{He}{ii}$ emission
flux and the $\ion{He}{ii}$/Ly$\alpha$ ratio across adjacent regions in the plane
of the sky (see Figs. \ref{FigHeII_OE} and \ref{FigLR})?
ii) what constraints can we derive on the gas density distribution
from the absolute values (or limit) of the $\ion{He}{ii}$/Ly$\alpha$ ratios?

We will start by examining the effect of limited spatial resolution
on the measured line emission ratios produced by two different ions
for a broad probability distribution function (PDF) of gas densities.  
We will then discriminate between different physical scenarios  
for the origin of both Ly$\alpha$ and $\ion{He}{ii}$ emission (or lack thereof)
and the large $\ion{He}{ii}$/Ly$\alpha$ ratio variations.
In particular, we will show that our results are best
explained by fluorescent recombination radiation produced by regions 
that are located about 1 Mpc from the quasar along our line of sight. 
Finally, we will show that at least the brightest part of the Slug 
should be associated with a very broad cold gas
density distribution that, if represented by a lognormal, would
imply dispersions as high as the one expected in the Interstellar
Medium (ISM) of galaxies \citep[see e.g.,][]{elmegreen02}.
Finally, we will put our result in the context of other giant Ly$\alpha$
nebulae discovered around type-I and type-II AGN (mostly radio-galaxies).

\subsection{Observed line ratios and gas density distribution}
\label{Disc_sub1}

In this section, we emphasise that, when the gas density distribution 
within the photometric and spectroscopic aperture is inhomogeneous (as expected), 
the ``observed'' line ratio (e.g., $\langle F_{\rm{HeII}}\rangle/\langle F_{\rm{Ly\alpha}}\rangle$ 
as defined below) can be very different than the average ``intrinsic'' line ratio 
(e.g., $\langle F_{\rm{HeII}}/F_{\rm{Ly\alpha}}\rangle$) 
that would result from the knowledge of local densities in every point in space.
In particular, this applies to all line emission that results from 
two-body processes (including, e.g., recombinations and collisional excitations)
because their emission scales as density squared.

For instance, the ``measured'' $\ion{He}{ii}$/Ly$\alpha$ line ratio produced by
recombination processes 
 { 
(in absence of dust and radiative transfer effects)
}
 is defined, from an observational point
of view, as:
\begin{equation}
\label{eq1}
\frac{<F_{\rm{HeII}}>}{<F_{\rm{Ly\alpha}}>}=\frac{h\nu_{\rm{HeII}}}{h\nu_{\rm{Ly\alpha}}}
\frac{\alpha_{\rm{HeII}}^{\rm{eff}}(T)}{\alpha_{\rm{Ly\alpha}}^{\rm{eff}}(T)}
\frac{<n_e n_{\rm{HeIII}}>}{<n_e n_p>} \ ,
\end{equation}
where the average (indicated by the symbols ``$ < > $'') is performed over the photometric 
and spectroscopic aperture or, analogously, within the spatial and spectral resolution element 
(and captures the idea that the flux is an integrated measurement). The 
temperature-dependent effective recombination coefficients for the $\ion{He}{ii}\lambda1640$ 
and Ly$\alpha$ line are indicated by $\alpha_{\rm{HeII}}^{\rm{eff}}$ and 
$\alpha_{\rm{Ly\alpha}}^{\rm{eff}}$, respectively
 {
\begin{footnote}{we use 
the following values of the effective recombination coefficients at 
$T=2\times10^{4}$K (Case A), from \citep{osterbrock}:
$\alpha_{\rm{Ly\alpha}}^{\rm{eff}}=9.1\times10^{-14}$ cm$^3$ s$^{-1}$ 
and $\alpha_{\rm{HeII}}^{\rm{eff}}=3.2\times10^{-13}$ cm$^3$ s$^{-1}$. 
The Case B coefficient value for Ly$\alpha$ is similar while the $\ion{He}{ii}$ coefficient
is higher by a factor of about 1.4 .}\end{footnote}
}
.
 In eq.\ref{eq1}, we have assumed
that the emitting gas within the photometric and spectroscopic aperture has a constant
temperature. This is a reasonable approximation for photoionized and metal poor gas 
in the low-density limit ($n<10^{4}$ cm$^{-3}$), if in thermal equilibrium
(e.g., \citealt{osterbrock}).
Substituting the following expressions that assume primordial helium 
abundance and neglecting the small contribution
of ionised helium to the electron density (up to a factor of about 
1.2):
\begin{equation}
\begin{split}
n_{\rm{HeIII}} & =0.087 n_{\rm{H}} x_{\rm{HeIII}} \ , \\
n_p & \equiv n_{\rm{H}} x_{\rm{HII}} \ , \\
n_e & \simeq n_{\rm{H}} x_{\rm{HII}} \ ,
\end{split}
\end{equation}
we obtain:
\begin{equation}
\frac{<F_{\rm{HeII}}>}{<F_{\rm{Ly\alpha}}>} \simeq R_0(T) 
\frac{<n_{\rm{H}}^2 x_{\rm{HeIII}} x_{\rm{HII}}>}{<n_{\rm{H}}^2 x_{\rm{HII}}^2>} \ ,
\label{eq_LR_full}
\end{equation}
where:
\begin{equation}
R_0(T) \equiv 0.087 \frac{\nu_{\rm{HeII}}\alpha_{\rm{HeII}}^{\rm{eff}}(T)}{\nu_{\rm{Ly\alpha}}\alpha_{\rm{Ly\alpha}}^{\rm{eff}}(T)}
 \ .
\end{equation}
Note that, for a temperature of $T=2\times10^{4}$K, $R_0\simeq0.23$ and $R_0\simeq0.3$ 
for Case A and Case B, respectively.

Equation \ref{eq_LR_full} can be simplified further assuming that the hydrogen is mostly
ionised (i.e., $x_{\rm{HII}}\simeq1$), as will typically be the case for the Slug nebula 
up to very high densities and large distances as we will show below, obtaining:
\begin{equation}
\begin{split}
\frac{<F_{\rm{HeII}}>}{<F_{\rm{Ly\alpha}}>} \simeq R_0(T) 
\frac{<n_{\rm{H}}^2 x_{\rm{HeIII}}>}{<n_{\rm{H}}^2>} \\
= R_0(T) \frac{\int_V x_{\rm{HeIII}} n_{\rm{H}}^2 dV}{\int_V n_{\rm{H}}^2 dV} \ ,
\end{split}
\label{eq_LR}
\end{equation}
where $V$ denotes the volume given by the photometric aperture (or spatial resolution element)
and the spectral integration window. 
The expression above can be rewritten in terms of the density distribution function $p(n)$ as:
\begin{equation}
\frac{<F_{\rm{HeII}}>}{<F_{\rm{Ly\alpha}}>} \simeq R_0(T) 
\frac{\int x_{\rm{HeIII}} n_{\rm{H}}^2 p(n_{\rm{H}}) d{n_{\rm{H}}}}
{\int n_{\rm{H}}^2 p(n_{\rm{H}}) d{n_{\rm{H}}}} \ ,
\end{equation}

As is clear from the expressions above, the ``measured'' 
$\ion{He}{ii}$/Ly$\alpha$ ratio for recombination radiation for highly ionised hydrogen gas
will scale with the average fraction of doubly ionised helium, $x_{\rm{HeIII}}$, 
\emph{weighted by the gas density squared}.
We note that $x_{\rm{HeIII}}$ is in general a function of density, incident flux above
4 Rydberg (i.e. ionization parameter) and temperature. However, at a given distance
from the quasar, the incident flux and temperature (due to photo-heating) will
be fixed or within a limited range and therefore $x_{\rm{HeIII}}$ 
would mainly depend on density. 

There is only one case in which the ``measured'' line ratio 
as defined above is equal to the  average ``intrinsic'' one
(e.g., $\langle F_{\rm{HeII}}/F_{\rm{Ly\alpha}}\rangle$),
that is when $p(n_{\rm{H}})$ is a delta function. For any other
density distribution, instead, the ``measured'' line ratio will be
always smaller than the ``intrinsic'' value 
because $x_{\rm{HeIII}}$ decreases at higher densities and
because of the $n_{\rm{H}}^2$ weighting.

When both hydrogen and helium are highly ionised, 
both the ``measured'' and ``intrinsic'' line ratios
will tend to the maximum value $R_0(T)$ that is
indeed independent of density.
It is interesting to note that our measured $\ion{He}{ii}$/Ly$\alpha$
ratio both in the ``region c'' ($\simeq0.05$) 
and the upper limit in the ``bright tail'' 
($\simeq 0.006$ at the 1$\sigma$ level) 
are significantly below $R_0(T)$ around temperatures 
of a few times $10^{4}$ K for both Case A ($\simeq0.23$) and 
Case B($\simeq0.3$). This is suggesting that helium cannot be
significantly doubly ionised
 {
(see also \citealt{fab15})
}
Moreover, as we will see below in detail, the ``measured'' line 
ratio in our case is low enough to provide a strong constraint on the 
clumpiness of the gas density distribution for the recombination
scenario\begin{footnote}{
 {
we stress that the results presented in this section apply to any recombination line ratio 
that involves two species that have very different critical densities as defined, 
e.g., in equations \ref{eq_ncrit_HI} and \ref{eq_ncrit_HeII} for hydrogen and single ionized helium,
respectively.
}
}\end{footnote}.

\subsection{On the origin of the large $\ion{He}{ii}$/Ly$\alpha$ gradient}
 
 In view of the discussion above, the possible origin of the
 strong ``measured'' line ratio variation across nearby spatial location
  within the Slug nebula include:
i) a variation in Ly$\alpha$ emission mechanism, e.g. recombination versus quasar broad-line-region 
scattering, 
ii) quasar emission variability (in time, opening angle and spectral properties),
iii) ionisation due to different sources than quasar ``a'',
iv) different density distribution, v) different physical distances. 

 The first possibility is readily excluded by the detection of H$\alpha$ emission
 from the ``bright tail'' of the Slug by \citet{leibler18}, i.e. from
 the same region where $\ion{He}{ii}$ is not detected and the measured $\ion{He}{ii}$/Ly$\alpha$
 upper limit is the lowest. In particular, the relatively large H$\alpha$ emission
 measured from this region exclude any significant contribution to the 
 Ly$\alpha$ emission from scattering of the quasar broad line regions photons.
 
 Another possibility is that the ``bright tail'' region without detected $\ion{He}{ii}$ emission
 does not receive a significant amount of photons above 4 Rydberg from quasar ``a'' due to, e.g. 
  {
 time variability effects (see e.g., \citealt{peterson04}, \citealt{vanden04}, \citealt{ross18} and references therein), quasar partial 
 obscuration (see e.g., \citealt{elvis00}, \citealt{dong05}, \citealt{gaskell18} and references therein) 
 or because of possible spectral ``hardness" variations along 
 different directions}
 \begin{footnote}{
  {
 this is easily illustrated in the case of a equal delta function density distribution 
 for both regions and in the high density regime (eq. 10) where the quotient of line ratios 
 is simply proportional to the ratios of $\Gamma$ as discussed at the end of this section. 
 A given ratio of the two $\Gamma_{\mathrm{HeII}}$ can be explained either as a distance effect (as we argue in this section), 
 or alternatively as a difference $\Delta_{\rm ion}$ in the slope of the ionizing spectrum as seen by different regions. 
 With all other parameters fixed, and assuming that the spectrum seeing by the ``region c" has the 
 standard slope ($\alpha=-1.7$) the ratio in eq. 10 would then roughly scale as $4^{-\Delta_{\mathrm{ion}}}\times4.7/(4.7+\Delta_{\rm ion})$. 
 }
 }
 \end{footnote}. Although this scenario would easily explain
 even a extremely low $\ion{He}{ii}$/Ly$\alpha$ ratio and strong spatial gradients,
 it would be very difficult to reconcile the fact that the line ratio variations seem to
 correlate extremely well with kinematical variations in terms of Ly$\alpha$ line 
 centroid (e.g., \citealt{leibler18}). 
 
 The presence of source ``c'' within the $\ion{He}{ii}$ detected region could hint at the
 possibility that different sources are responsible for the ionisation of different
 part of the nebula, particularly if source ``c'' harbours an Active Galactic Nucleus (AGN).
 If this source were fully ionizing both hydrogen and helium, we would have
 expected to see a line ratio approaching 0.3 (Case B) or 0.23 (Case A)
 as discussed in section 4.1. However, the measured line ratio is much
 below these values. Therefore, if source ``c" is responsible for the
 photoionization of ``region c" one would have expected to see
 variations in the $\ion{He}{ii}$/Ly$\alpha$ ratio close
 to the location of this source. This is because ionisation effects should scale as
 $1/r^2$ (see below for details). However, as shown in Fig.\ref{FigLR}, the line ratio is rather 
 constant around the location of source ``c''. 
 This would require a fine tuned variation in the gas density distribution
 to balance the varying flux in order to produce the absence of line ratio variations
 across the location of source ``c''. We consider this possibility unlikely.
 Moreover, both from the infrared observation of \citet{leibler18} and from
 the narrowness of the rest-frame UV emission lines it is very unlikely that
 source ``c'' could harbor an AGN bright enough to produce both the extended $\ion{He}{ii}$ and
 Ly$\alpha$ emission (the same applies considering the relatively low SFR 
 of this sources derived in the previous sections).
 The most likely hypothesis therefore is that the 4 Rydberg ``illumination'' is coming
 from the more distant but much brighter quasar ``a''. 
  Similarly, the absence of detectable bright continuum sources in the ``bright tail'' region
 (see Fig.\ref{FigRGB}) suggests that ultra-luminous quasar ``a'' is the most likely source of 
 ``illumination'' for this region.
  {
 The only other securely detected AGN in this field, the quasar companion ``b", is more than 5 magnitudes
 fainter than quasar ``a" and even more distant in projected space (although there is 
 large uncertainty in redshift for this quasar) from both ``region c" and the ``bright tail"
 with respect to the other possible sources considered here.
 Finally, we notice that including any possible additional contribution 
 to the helium ionising flux from quasar ``b" or even source ``c" 
 with respect to quasar ``a" would strengthen the requirement
 for large gas densities as discussed below and in section \ref{secOrigin}.
 }

 By excluding the scenarios above as the least plausible we are left with
 the possibilities that the line ratio variations are due to either 
 gas density distribution variations (as discussed in \ref{Disc_sub1}) 
 or different physical distances, or both. 
 On this regard, it is important to notice that the gradient in the 
 $\ion{He}{ii}$/Ly$\alpha$ ratio is mostly driven by a strong variation in the $\ion{He}{ii}$ 
 emission. Indeed the Ly$\alpha$ SB of the ``region c'' and ``bright tail''
 are very similar. In the plausible assumption that the hydrogen is highly
 ionised in both regions, as we will demonstrate later, any density variation
 across the two regions should produce a significant difference in
 Ly$\alpha$ SB. For instance, in the highly simplified case in which 
 the emitting gas density distribution is constant, 
 the Ly$\alpha$ emission from recombination radiation would scale as the gas density squared 
 while the line ratio would only scale about linearly with density, as discussed below.
 In more general cases, discussed in the next section, we will show that
 indeed the Ly$\alpha$ SB is more sensitive to density variation than
 the line ratio. 
 
 The most likely hypothesis therefore is that different physical distances
 of the two regions from the quasar produce 
 the lack of detectable $\ion{He}{ii}$ emission that results in the strong 
 observed gradient in the $\ion{He}{ii}$/Ly$\alpha$ ratio.
 This suggestion is reinforced by the fact that the $\ion{He}{ii}$/Ly$\alpha$ gradient 
 arises exactly at the spatial location where a strong and abrupt Ly$\alpha$ 
 velocity shift is present (see e.g., \citealt{leibler18})
 In particular, the velocity shift between the ``bright tail'' and ``region c''
 is as large as 900 km/s as measured from Ly$\alpha$, H$\alpha$
 and $\ion{He}{ii}$ emission. This is much larger than
 the virial velocity of a dark matter halo with mass of about 
 $10^{13}$  solar masses at this redshift (about 450 km/s). 
 If completely due to Hubble flow, this velocity shift
 would correspond to physical distances as large as 4 Mpc. 
 Note that the quasar ``a'' systemic redshift is located in between 
 these two regions (-350 km/s from ``region c'' and +650 km/s
 from the ``bright tail'').
  {However, because peculiar velocities as large as a few hundreds
 of km/s are expected in such an environment, it is difficult
 to firmly establish if the quasar is physically between these
 two regions along our line of sight or in the background. 
 }
 
 In the next section, we will evaluate in detail the expected
 line ratios for a given density distribution function and distance
 from the quasar. However, it is instructive here to consider the simplest
 case in which the emitting gas density distribution is constant
 (i.e. is a delta function $p(n)=\delta(n-n_0)$) 
  {and equal}
 for both regions. In this case, we can simply evaluate in which
 situations the different line ratios could be explained just
 in terms of different relative distances from the quasar.
 Assuming once again that hydrogen is highly ionised (implying
 both $x_{\rm{HI}}\simeq 0$ and $x_{\rm{HeI}}\simeq 0$), 
 it is easy to show that:
 \begin{equation}
 x_{\rm{HeIII}}=\frac{\Gamma_{\rm{HeII}}}{\Gamma_{\rm{HeII}}+n_0\alpha_{\rm{HeIII}}} \ ,
 \end{equation}
 and, therefore using eq. \ref{eq_LR} that:
 \begin{equation}
 \frac{LR^c}{LR^{tail}}\simeq \frac{\Gamma_{\rm{HeII}}^c}{\Gamma_{\rm{HeII}}^{tail}}
\times \left( \frac{\Gamma_{\rm{HeII}}^{tail}+n_0\alpha_{\rm{HeIII}}}{\Gamma_{\rm{HeII}}^{c}+n_0\alpha_{\rm{HeIII}}}\right) \ ,
\label{eq_comp}
\end{equation}
 where $LR^c$ and $LR^{tail}$ represent the measured line ratio in ``region c'' and the ``bright tail'', respectively,
 while $\Gamma_{\rm{HeII}}^c$ and $\Gamma_{\rm{HeII}}^{tail}$ are the corresponding $\ion{He}{ii}$ photoionisation
 rates in these regions. Finally, $\alpha_{\rm{HeIII}}$ denotes the temperature dependent $\ion{He}{ii}$I recombination
 coefficient for which we use a value of $1.3\times10^{-12}$ cm$^{3}$ s$^{-1}$ at $T\sim2\times10^{4}$ K. 
 Given the observed continuum luminosity of our quasar and 
 a typical spectral profile in the extreme UV as in \citet{lusso15} the $\ion{He}{ii}$ photoionization rate is given by:
 \begin{equation}
 \Gamma_{\rm{HeII}}\simeq 9.2 \times 10^{-12} \left( \frac{r}{500 \rm{kpc}}\right)^{-2} \rm{s^{-1}} \ ,
 \end{equation}
 where $r$ denotes the physical distance between the quasar and the gas cloud. 
 When $\Gamma_{\rm{HeII}}^c/(n_0\alpha_{\rm{HeIII}})\ll1$ (and similarly for the tail region), 
 corresponding to, e.g.,
 $n_0>7$ cm$^{-3}$ at $r=500$ kpc, equation \ref{eq_comp} can be approximated as:
  \begin{equation}
 \frac{LR^c}{LR^{tail}}\simeq \frac{\Gamma_{\rm{HeII}}^c}{\Gamma_{\rm{HeII}}^{tail}} \ ,
 \end{equation}
 implying that a gradient of about a factor of ten in the line ratio could be easily explained,
 in this simplified case, if the ``bright tail'' region is about three times more distant
 than the ``region c'' with respect to the quasar. For smaller values of $n_0$ this
 ratio of distances increases to a factor of about four when 
 $\Gamma_{\rm{HeII}}^c/(n_0\alpha_{\rm{HeIII}})\sim1$.
 In case a broad density distribution is used, the required ratio in relative distances
 can be again reduced to about a factor of three, even if the average density is 
 much below the values discussed above, as we will see in the next section. 
 It is interesting to note that this factor of three is totally consistent with the
 kinematical constraints discussed above.
 
 Using similar arguments as before, it is simple to verify that if the two regions 
 are placed at the same distance (and therefore they have the same 
 $\Gamma_{\rm{HeII}}$), a factor of ten variation in the line ratio would imply
 a density ratio at least as high as this (assuming that the density distributions are
 delta functions). 
  As mentioned above, this would therefore imply a 
 change in the Ly$\alpha$ SB by a factor $n_0^{2}$, i.e. by a factor
 of at least 100, which is indeed not observed.
 
 In this section, we have assumed that the hydrogen is 
 mostly ionised. 
 This is a reasonable assumption because the density values at which
 hydrogen becomes neutral are very large, given the expected large
 value of the hydrogen photoionisation rate for UM287 (obtained as above)
  {
 in the conservative assumption that this is the only source of ionisation:
 }
 \begin{equation}
 \Gamma_{\rm{HI}}\simeq 3.9 \times 10^{-10} \left( \frac{r}{500\ \rm{kpc}}\right)^{-2} \rm{s^{-1}} \ .
 \end{equation}
 Indeed, assuming a temperature of $2\times10^{4}$ K and the case A recombination 
 coefficient $\alpha_{\rm{HII}}\simeq2.5\times10^{-13}$ cm$^{3}$ s$^{-1}$, 
 the hydrogen will become mostly neutral above the following density:
 \begin{equation}\label{eq_ncrit_HI}
 n_{\mathrm{H}}^{\mathrm{HI,crit}}\simeq \frac{\Gamma_{\rm{HI}}}{\alpha_{\rm{HII}}}\simeq 1500 \left( \frac{r}{500\ \rm{kpc}}\right)^{-2} \rm{cm^{-3}} .
 \end{equation}
 As a comparison, the density for which $\ion{He}{iii}$ becomes $\ion{He}{ii}$, as derived above, is about 200 times smaller:
 \begin{equation}\label{eq_ncrit_HeII}
 n_{\mathrm{H}}^{\rm{HeII,crit}}\simeq \frac{\Gamma_{\rm{HeII}}}{\alpha_{\rm{HeIII}}}\simeq 7 \left( \frac{r}{500\ \rm{kpc}}\right)^{-2} \rm{cm^{-3}} .
 \end{equation}
 There is therefore a large range of densities at which hydrogen is still ionised 
 while most of the doubly ionised helium is not present. 
 In the next section, we will show the result of our full calculation that
 takes into account the proper ionised fraction at each density.

 \subsection{On the origin of the small $\ion{He}{ii}$/Ly$\alpha$ values}\label{secOrigin}
 
 In the previous section, we have discussed how the strong gradients
 in the $\ion{He}{ii}$/Ly$\alpha$ ratio combined with kinematic information
 and the presence of H$\alpha$ emission, suggest that the ``bright tail''
 regions should be at least three times more distant from the quasar ``a''
 than ``region c'' (in the case of constant emitting gas density distribution). 
 In this section, we explore which constraints
 on the (unresolved) emitting gas density distribution 
 and absolute distances can be derived from the 
 measured values (or limits) of the $\ion{He}{ii}$/Ly$\alpha$ ratios. 
 As in the previous section, we will make the plausible assumption that the main emission
 mechanism for both lines is recombination radiation and that scattering from 
 the quasar broad line region is negligible (as implied by the detection of H$\alpha$
 emission). Collisional excitation can be excluded for the $\ion{He}{ii}\lambda1640$ line, 
 as it would require electron temperatures of about 10$^{5}$ K  that are
 difficult to produce for photo-ionised and dense gas, even for a quasar
 spectrum (that would range between $2-5\times10^{4}$ K, e.g. \citealt{sc08}).
  Collisionally-excited Ly$\alpha$ emission could be produced 
 instead efficiently at the expected temperatures (e.g. \citealt{sc08})
 but the volume occupied by partially ionised dense gas, if present at all, will 
 be negligible with respect to the ionized volume 
 (see section 4.2).
  {
 Finally, we will make the conservative assumption that
 quasar ``a" is the only source of ionisation.
 }
 
 
 \subsubsection{Maximum distance from quasar ``a''}
 
 We have shown in section \ref{Disc_sub1} that the ``measured'' 
 line ratio can be very sensitive to the emitting gas density distribution within 
 the photometric and spectroscopic aperture.
 In particular, we expect that a broader density distribution function
 at a fixed average density will produce lower line ratios.  
 Any constraint on the density distribution would be however
 degenerate with the value of the photoionisation rate of $\ion{He}{ii}$,
 that, in turn depends on the distance of the cloud. 
 In particular, we expect that at larger distances, smaller densities
 would be required to produce a low line ratio.
 It is therefore important to derive some independent constraints
 on, e.g., the maximum distance at which the ``bright tail'' region
 could be placed, in order to derive meaningful constraints on its
 gas density distribution from the $\ion{He}{ii}$/Ly$\alpha$ ratio.
 
 Such constraints could be derived by the self-shielding limit
 for the Ly$\alpha$ fluorescent surface brightness produced
 by quasar ``a'' (e.g., \citealt{sc05}). In this limit, reached when
 the total optical depth to hydrogen ionising photons 
 becomes much larger than one, the expected emission is
 independent of local densities and depends only
 on the impinging ionising flux.
 In particular, using the observed luminosity of quasar ``a'' (UM287) and assuming
 the same spectrum as in the previous section, the maximum
 distance as a function of the observed Ly$\alpha$ SB will be 
 (see also \citealt{fab15}):
 \begin{equation}
 r_{\rm{max}} \simeq  1\ \rm{Mpc} \times
 \left( \frac{ \rm{SB_{Ly\alpha, 17}} }{2.25} \right)^{-0.5}\left( \frac{f_{\rm{C}}}{1.0}\right)^{0.5} 
 \left(\frac{\Gamma^{\rm{act}}_{\rm{HI}}}{\Gamma_{\rm{HI}}^{\rm{obs}}}\right)^{0.5} 
 \end{equation}
  where $\rm{SB_{Ly\alpha, 17}}$ is the observed Ly$\alpha$ SB in units of 
  $10^{-17}\rm{erg\ s^{-1}\ cm^{-2}\ arcsec^2}$, $f_{\rm{C}}$ is the self-shielded
  gas covering fraction within the spatial resolution element, 
  $\Gamma_{\rm{HI}}^{\rm{obs}}$ is the inferred photoionisation rate for UM287 using the currently 
  observed quasar luminosity (along our line of sight), and $\Gamma^{\rm{act}}_{\rm{HI}}$ 
  is the actual photoionization rate at the location of the optically thick gas. 
  Note that both $f_{\rm{C}}$ and $\Gamma^{\rm{act}}_{\rm{HI}}$ could be uncertain 
  within a factor of a few.
  
 The observed Ly$\alpha$ SB in both the ``bright tail'' and ``region c'' is around 
 $2.5\times10^{-17}\rm{erg\ s^{-1}\ cm^{-2}\ arcsec^2}$ corresponding to a maximum
 distance of about 1 physical Mpc. This distance would be larger if the observed SB 
 is decreased because of local radiative transfer effects or absorption along our line of sight.
 For similar reasons, the quoted Ly$\alpha$ SBs in the reminder of this section
 should be considered as upper limits. 
  {
 We also note that there is very little or no spatial overlap in the Ly$\alpha$ image
 between the ``bright tail" and ``region c" as they are very well separated 
 in velocity space without signatures of double peaked emission (\citealt{leibler18}).
 }

 \subsubsection{Delta function density distribution}
 
 Before moving to more general density distributions, it is interesting to consider again 
 the extremely simplified case of the delta function $p(n)=\delta(n_0)$ 
 and to derive the minimum densities needed to explain the $\ion{He}{ii}$/Ly$\alpha$ upper
 limits in the ``bright tail'' if placed at the maximum distance of 1 Mpc. 
 Using the results of the previous section, a temperature of $T=2\times10^4$K, 
 and assuming conservatively the 2$\sigma$ upper limit of 0.012 for the 
 $\ion{He}{ii}$/Ly$\alpha$ ratio we derive a density of $n_0\simeq30$ cm$^{-3}$
 for Case A and  $n_0\simeq75$ cm$^{-3}$ for Case B (for both hydrogen and helium). 
 As shown in the previous section, these densities would also explain 
 the measured line ratio in ``region c'' if located at a distance 
 of about 300 kpc from quasar ``a''. The derived densities 
 increase as the square root of the distance from the quasar
 and the values quoted above should be considered as an absolute
 minimum for a delta function density distribution of the (cold) emitting gas. 
 Such high densities, combined with the observed Ly$\alpha$ SB would
 imply an extremely small volume filling factor of the order of $f_V\simeq10^{-6}$,
 if each of the two regions has a thickness along our line of sight of 
 about 100 kpc (see, e.g. equation 3 in \citealt{sc17}
  {
 \begin{footnote}{this equation does not explicitly contains $f_V$ (assumed to be one)
 but it can be simply rewritten including this factor considering that the
 Ly$\alpha$ SB scales linearly with $f_V$.}\end{footnote}
 }
 ). 
 
 Unless these clouds are gravitationally bound, 
 we expect that such high densities would be quickly dismantled in a short timescale:
 these clouds cannot be pressure confined because the hot gas surrounding them 
 should have temperatures or densities that are at least one order of magnitude larger than what structure
 formation could reasonably provide. For instance, the virial temperature and densities
 of a 10$^{13}$ M$_{\odot}$ dark matter halo at $z\simeq2.3$ are expected to be
 around $3\times10^7$ K and 10$^{-3}$ cm$^{-3}$, respectively. 
 Therefore, even in the very unlikely hypothesis that both ''region c'' and the ''bright tail region'' 
 are associated with such massive haloes, only gas clouds with densities of about
 1.5 cm$^{-3}$ could be pressure confined. In order to confine gas clouds 
 with a density of $n_0\simeq30$ cm$^{-3}$ once photoionized by the quasar
 (and therefore at a temperature of about 2$\times10^{4}$ K), 
 we would require either a hot gas temperature of $6\times10^8$ K or a hot
 gas density that is 20 times higher than the virial density. 
 Alternatively, the temperature of the cold clouds should be initially much 
 lower than 10$^{3}$ K, implying that these clouds are in the process of 
 photo-evaporating after being illuminated by the quasar. 
 All these situation are problematic because they either require extreme
 properties for the confining hot gas or that the cold clouds are extremely
 short lived, with obvious implication for the observability of giant Ly$\alpha$ nebulae.
 
 \subsubsection{Log-normal density distribution}\label{subsubLN}
 
 These problems can be solved by relaxing one of the extreme simplifications
made above (and in general in other photo-ionisation models in the literature),
i.e. that the emitting gas density distribution is a delta function. 
As demonstrated in section \ref{Disc_sub1}, a broad density distribution may 
decrease the ``observed'' line ratio by a large factor while keeping 
the same volume-averaged density. 
Broad density distributions are commonly observed 
in multiphase media like, e.g., the ISM of our galaxy 
\citep[e.g.][]{myers78}.
In particular, both simulations and observations suggest that
gas densities in a globally stable and turbulent ISM is well fitted 
by a lognormal Probability Distribution Function (PDF)
(e.g, \citealt{wada07} and references therein):
\begin{equation}
\mathrm{PDF}(n)dn = \frac{1}{\sqrt{2\pi}\sigma} \exp{ \left[ - \frac{[\mathrm{ln}(n/n_1)]^2}{2\sigma^2}\right] } d\mathrm{ln}(n),
\end{equation} 
where $\sigma$ is the lognormal dispersion and $n_1$ is a characteristic density that is 
connected to the average volume density by the relation:
\begin{equation}
\langle n \rangle_c=n_1 \exp{\frac{\sigma^2}{2}} .
\end{equation}
%
Numerical studies suggested that a lognormal distribution is
characteristic of isothermal, turbulent flow and that $\sigma$ 
is determined by the ``one-dimensional Mach number ($M$)'' of the turbulent
motion following the relation: $e^{\sigma^2}\sim1+3M^2/4$ \citep[e.g.,][]{padoan02}. 
Although a discussion of the origin of the gas density distribution in the ISM
and its effect on the galactic star formation is clearly beyond the scope of this paper, 
we notice that a large value of $\sigma$ (e.g., $\sigma\sim2.3$) has been suggested 
as a key requirement to reproduce the Schmidt-Kennicutt law \citep[see e.g.,][]{elmegreen02,wada07}. 

In the remainder of this section, we use a lognormal density distribution as a first possible ansatz 
for the PDF of the emitting gas in the Slug nebula, with the assumption that some of the
processes that may be responsible for the appearance of such a PDF in the ISM
may be operating also in our case. Our main requirement is that the cold emitting gas is
``on average'' in pressure equilibrium with a hot confining medium, 
i.e. that the density averaged over the volume of the cold gas ($\langle n \rangle_c$) should be determined by
the temperature and density of the confining hot gas. In this context, the broadness
of the gas density distribution represents a perturbation on densities that could be caused
by, e.g., turbulence, sounds waves, or other (unknown) processes that
may act on both local or large scales. Examples of large-scale perturbation may be caused by
gravitational accretion combined with hydrodynamical (e.g., Kelvin-Helmholtz) or thermal instabilities 
as we will discuss elsewhere (Vossberg, Cantalupo \& Pezzulli, in prep.; see also \citealt{mandelker16,padnos18,ji18}). 
Because these density perturbations will be either highly over-pressured or under-pressured for large 
$\sigma$ we expect that they would dissipate quickly. Therefore, we expect that $\sigma$
will be reduced over time if the perturbation mechanism is not acting continuously or if the resulting structures
do not become self-gravitating.

Our working hypothesis is that the confining hot gas is virialized within 
dark matter haloes that are part of the cosmic web around quasar ``a''.
The hot gas density is therefore fixed at about 200 times the average density of the universe at
$z\simeq2.3$, i.e., $n_{\rm{hot}}\simeq1.4\times10^{-3}$ cm$^{-3}$ and the temperature
is assumed equal to the virial temperature of a given dark matter halo mass at this redshift. 
In turn, this temperature is related to the average density of the cold gas (assumed to be
at a temperature of $2\times10^4$ K) by the pressure-confinement condition discussed above. 
The other ingredients of our simple model are: i) the mass fraction of gas in the cold component 
within the virial radius ($f_{\rm{cold}}$) which we assume to be 10\%, 
and ii) the size of the emitting region along the line of sight that we assume to be 100 kpc. 
 Given the densities of the hot and cold components, this
cold mass fraction translates into a given volume filling factor that determines the expected
Ly$\alpha$ SB. We stress that the actual value of $f_{\rm{cold}}$ only affects
the Ly$\alpha$ SB and that any increase in $f_{\rm{cold}}$ can be compensated by a smaller
size of the emitting region along the line of sight, producing the same results. 

The running parameters in the model are therefore: i) the average volume density
of the cold gas component $\langle n \rangle_c$, ii) the log-normal dispersion $\sigma$, and iii) the
distances of ``region c'' and the ``bright tail'' from quasar ``a''. 
These four parameters will be fixed by the observed Ly$\alpha$ SB and the $\ion{He}{ii}$/Ly$\alpha$ 
line ratios of both regions, under the assumption that they are characterised by the same
density distributions. Given the assumption of ``average'' pressure equilibrium above, 
the masses of the hosting dark matter haloes of these structures will be directly linked
to the average volume density of the cold gas $\langle n \rangle_c$. 

\begin{figure}
\includegraphics[width=1.03\columnwidth]{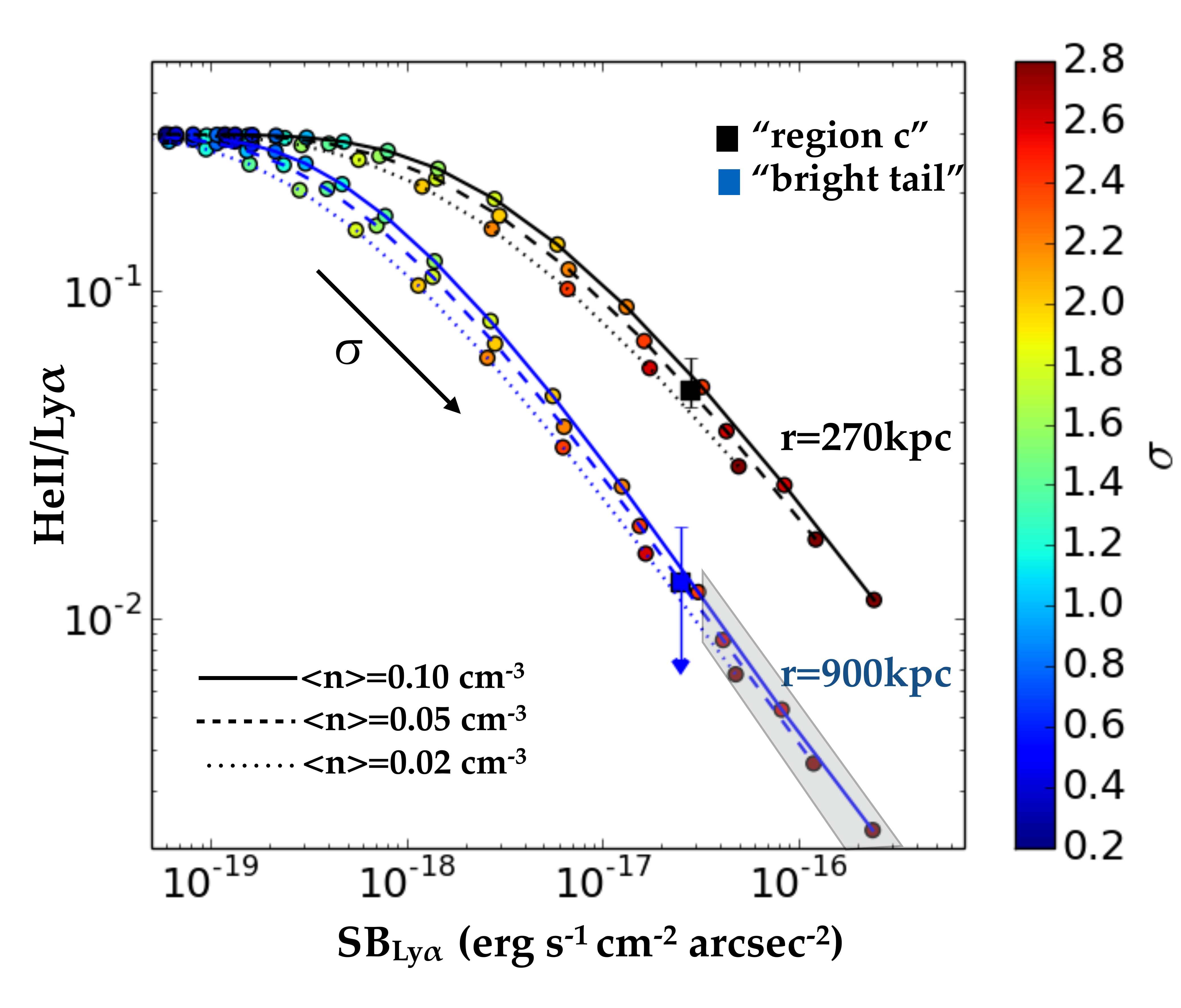}
\caption{
Results of our photoionisation models
for case B recombination emission (lines) compared to the observed 
Ly$\alpha$ SB and He/Ly$\alpha$ ratios
for both ``region c'' (black square; 
 {
error bars include the maximum effect of 
the quasar Ly$\alpha$ PSF, see section \ref{secLineratio}) 
and ``bright tail'' 
(blue square, indicating the 2$\sigma$ upper limit; the upper error bar 
indicates the 3$\sigma$ upper limit; 
for simplicity, we use the 2$\sigma$ upper limit as a measured point for the line ratio),
in the conservative assumption that quasar ``a" is the only source of ionisation.
}
The predicted values have been obtained from eq. 1
by numerically solving the combined photoinization equilibrium equations 
for both hydrogen and helium in all their possible ionisation states at each density 
given by a lognormal distribution with dispersion $\sigma$ and average volume 
density $\langle n \rangle_c$. Two sets of lines are shown depending on the cloud distance
from quasar ``a'':  900kpc (blue)
and 270kpc (black). 
For each set of lines we run our models with three different $\langle n \rangle_c$:
i) $\langle n \rangle_c=0.02$ cm$^{-3}$, corresponding to a dark matter host halo M$_h$ of about 10$^{11}$
M$_{\odot}$ (and volume filling factor of the cold emitting gas $f_v\simeq8\times10^{-3}$), 
ii) $\langle n \rangle_c=0.05$ cm$^{-3}$ (M$_h\simeq10^{11.5}$ M$_{\odot}$ and $f_v\simeq3\times10^{-3}$ ), and 
iii) $\langle n \rangle_c=0.1$ cm$^{-3}$ (M$_h\simeq10^{12.0}$ M$_{\odot}$ and $f_v\simeq10^{-3}$ ). 
Finally, each point in the lines represent a different value of $\sigma$ 
and we overlay several coloured circle equally spaced with $\Delta\sigma=0.2$
at $\sigma$ values indicated by the color bar. 
The regions of parameter space that are not allowed because of the 
hydrogen self-shielding limit are shaded in grey.
The models reproduce the observed values (or limits) only for 
broad density distribution with $2.4<\sigma<2.7$ and only if the 
gas associated with the ``bright tail'' is located at large distances from the quasar ``a'',
close to the self-shielding limit of about 1 Mpc. 
 {
We notice that, including any possible contribution to both hydrogen and helium
ionizing radiation from other sources would require even broader gas
density distributions and would move the self-shielding limit to even
larger SB.
}
See text for further discussion.
}

\label{Fig_LN_LR}
\end{figure}

In Fig. \ref{Fig_LN_LR}, we show the result of our photoionisation models
for case B recombination emission (lines) compared to the observed 
Ly$\alpha$ SB and He/Ly$\alpha$ ratios
for both the ``region c'' (black square;
 {
error bars include the maximum effect of 
the quasar Ly$\alpha$ PSF, see section \ref{secLineratio}) 
and ``bright tail'' 
(blue square, indicating the 2$\sigma$ upper limit; the upper error bar 
indicates the 3$\sigma$ upper limit; 
for simplicity, we use the 2$\sigma$ upper limit as a measured point for the line ratio).
}
The predicted values have been obtained from eq. 1
by numerically solving the combined photoionization equilibrium equations 
for both hydrogen and helium in all their possible ionisation states\begin{footnote}{
 {
see, e.g., equations 7, 8 and 9 in \citealt{radamesh} with time derivatives set to zero
and with $C$ set to 1.
}
}\end{footnote} at each density 
given by a lognormal distribution with dispersion $\sigma$ and average volume 
density $\langle n \rangle_c$. In particular, we show two sets of lines: the blue lines correspond
to a distance of 900 kpc from quasar ``a'' (slightly below the maximum distance 
for hydrogen self-shielding) and the black lines correspond to a distance of 270kpc.
For each set of lines we run our models with three different $\langle n \rangle_c$:
i) $\langle n \rangle_c=0.02$ cm$^{-3}$, corresponding to a dark matter host halo M$_h$ of about 10$^{11}$
M$_{\odot}$ (and volume filling factor of the cold emitting gas $f_v\simeq8\times10^{-3}$), 
ii) $\langle n \rangle_c=0.05$ cm$^{-3}$ (M$_h\simeq10^{11.5}$ M$_{\odot}$ and $f_v\simeq3\times10^{-3}$ ), and 
iii) $\langle n \rangle_c=0.1$ cm$^{-3}$ (M$_h\simeq10^{12.0}$ M$_{\odot}$ and $f_v\simeq10^{-3}$ ). 
These average densities span a range of plausible halo masses for the host haloes. 
Finally, each point along the lines represent a different value of $\sigma$ 
and we overlay, for clarity, several coloured circles equally spaced with $\Delta\sigma=0.2$
at $\sigma$ values indicated by the color bar. 
The regions of parameter space that are not allowed because of the 
hydrogen self-shielding limit are shaded in grey.
We note that the predicted Ly$\alpha$ SB scales linearly with the product of 
$f_v$ (determined by $\langle n \rangle_c$ and the cold mass fraction that we fixed to 10\%)
and the size of the emitting region along the line of sight, that we have assumed to be 100 kpc. 
However, the line ratio is of course independent of these parameters. 

As clear from the figure and as expected from our discussion of the ``observed''
versus ``intrinsic'' line ratios, the broadness of the density distribution is the
main factor driving the predicted line ratios to low values. 
In particular, no model is able to produce line ratios below 10\% unless $\sigma>1.5$.
Furthermore, large distances (close to the self-shielding limit for Case B recombination) 
are also required to explain the $\ion{He}{ii}$/Ly$\alpha$ limit obtained for the ``bright tail''.
Line ratios are instead less sensitive to the value of $\langle n \rangle_c$. However, at a
fixed distance, $\sigma$ and $\langle n \rangle_c$ are degenerate, in the sense that 
the observed line ratios and Ly$\alpha$ SB could be explained by a
smaller $\sigma$ if $\langle n \rangle_c$ is higher. For instance, if the ``bright tail''
is at 900 kpc from quasar ``a'', the upper limit on the line ratio and the
measured Ly$\alpha$ SB could be explained by a lognormal density
distribution with $2.4<\sigma<2.7$ for $(0.10\ \mathrm{cm^{-3}})> \langle n \rangle_c > (0.02\ \mathrm{cm^{-3}})$. 
As a reference, such values of $\sigma$ correspond to ``internal" clumping factors
($C_c\equiv\langle n^2 \rangle_c /\langle n \rangle_c^2$,  
that for a lognormal distribution is simply given by $C=\exp{(\sigma^2)}$) 
$300\lesssim C_c\lesssim 1000$ for the cold component,
in agreement with the estimated based on comparison with cosmological
simulations for the Ly$\alpha$ SB discussed in \citet{sc14}.
These values would also explain the measured line ratios and Ly$\alpha$
SB of the ``region c'' if placed at a distance of 270 kpc. 
 {
As expected, the $\ion{He}{ii}$/Ly$\alpha$ ratio and the Ly$\alpha$ SB are generally
anticorrelated. In particular, at a fixed average density $\langle n \rangle_c$,
 the Ly$\alpha$ SB increases linearly with the "internal" clumping factor
 $C=\exp{(\sigma^2)}$.
 At the same time, larger values of $\sigma$ produce smaller $\ion{He}{ii}$/Ly$\alpha$
 ratios. This means that, if the Ly$\alpha$ SB in some region of Ly$\alpha$ nebulae 
 is driven by large clumping factors then we necessarily expect low 
 $\ion{He}{ii}$/Ly$\alpha$ within the same regions. 
}

 {
We note that performing the calculation using the Case A effective and total recombination
coefficients would require slighter smaller 
values of $\sigma$ at a fixed distance to obtain the same $\ion{He}{ii}$/Ly$\alpha$ ratios,
i.e. $\sigma\simeq2.3$ instead of $\sigma\simeq2.5$ for the ``bright tail" for 
$\langle n \rangle_c=0.05$ cm$^{-3}$ (see Appendix).
On the other hand, at a fixed $\sigma=2.5$ and $\langle n \rangle_c=0.05$ cm$^{-3}$, 
the Case A calculation produces $\ion{He}{ii}$/Ly$\alpha$ ratios consistent 
with the $2\sigma$ upper limit of the ``bright tail" for distances equal or larger 
than about 550 kpc instead of 900 kpc.
For ``region c" 
the required distance decreases to about 180 kpc (see Appendix).
In all cases, however, a log-normal distribution 
with $\sigma>2.3$ is required to match the low $\ion{He}{ii}$/Ly$\alpha$ ratios,
at least for the ``bright tail".
}

As discussed above, the high values of $\sigma$ implied by our results are not dissimilar 
to the ones obtained for the ISM although the average density are at least an order of magnitude smaller.
 A detailed discussion of the possible origin of such broad density distributions 
 is beyond the scope of the current paper and will be the subject of future theoretical 
 studies (e.g., Vossberg, Cantalupo \& Pezzulli, in prep.).
The goal of the current analysis is to show that both large gradients 
and very low values of the observed $\ion{He}{ii}$/Ly$\alpha$ ratio can be produced by
log-normal density distributions with average densities
that are consistent with simple assumptions about pressure confinement.
We notice that a lognormal density distribution with $\sigma=2.5$ and
$\langle n \rangle_c=0.05$ cm$^{-3}$ still implies that a non-zero fraction of the volume
should be occupied by gas with densities similar or larger than
the value derived in the case of a delta-function PDF, i.e. $n\simeq75$ cm$^{-3}$.
However, the implied volume filling factor for such dense gas in the log-normal
case is $f_v (n>75$ cm$^{-3})\simeq5\times10^{-8}$, which is 
much smaller than the value obtained for the delta-function case discussed above. 

 {
Finally, we note that any possible contribution to both hydrogen and helium
ionizing radiation from other sources (e.g., the faint quasar ``b" or source ``c",
if an AGN is present) would require even broader gas
density distributions and would move the self-shielding limit to even
larger Ly$\alpha$ SB. This is because any increase in the photoionization rate, 
at a given spatial location, would necessarily require higher densities 
to produce the same ionisation state. Our choice of including only quasar
``a" as a possible ionisation source should therefore be regarded
as conservative for our main conclusions.
}

\subsection{Comparison to other giant Ly$\alpha$ nebulae: type-II versus type-I AGN}

%

Other giant Ly$\alpha$ nebulae, such as the one detected around
high redshift radio galaxies (see e.g., \citealt{miley08} and \citealt{villarmartin07rev}
for reviews) also show extended Ly$\alpha$ emission
that is in many cases associated with $\ion{He}{ii}$ emission. 
Integrated $\ion{He}{ii}$/Ly$\alpha$ ratios, including also the emission from the
radio-galaxy itself, are typically around 0.12 between $2<z<3$ 
\citep[e.g.,][]{villarmartin07}, i.e. much larger than our limit on the ``bright tail''.
However, line ratios measured from parts of the extended haloes
reach values as small as 0.03 \citep[e.g.,][]{villarmartin07}. 
Such low values have also been interpreted by previous authors
as either evidence for extremely low ionisation parameters and therefore 
high densities ($n\gg100$ cm$^{-3}$, 
once again in a delta-function density PDF scenario, 
see, e.g. \citealt{villarmartin07rev} and \citealt{humphrey18} for 
a recent example;
but see also \citealt{binette96} and \citealt{humphrey08}
) 
or even evidence for stellar photo-ionisation rather 
than AGN photo-ionisation 
 {
(\citealt{villarmartin07,hatch08,emonts18}
}
; this would
correspond to a decreased $\Gamma_{\rm{HeII}}$ with respect to
$\Gamma_{\rm{HI}}$). The latter hypothesis can be firmly
excluded in our case (and for the large majority of the giant Ly$\alpha$
nebulae discovered so far), since the rest-frame 
Ly$\alpha$ Equivalent Width (EW$_0$) of the ``bright tail'' is extremely large, 
i.e. EW$_0>3000\mathrm{\AA}$ (at $3\sigma$ confidence level considering an aperture
with diameter 3 arcsec centered on the ``bright tail'' region).
Such high EW cannot be produced by embedded star
formation (see e.g., \citealt{sc07} and \citealt{sc12} for discussions).

It is interesting to note that radio-loud Ly$\alpha$ haloes
have much broader kinematics with respect to radio-quiet systems
and that both the luminosity and kinematics of Ly$\alpha$ emission
seems to be associated with the presence of radio-jets 
 {although 
more quiescent kinematics are also present}
\citep{villarmartin07}.
In view of our discussion above, a broad density PDF could produce
such low $\ion{He}{ii}$/Ly$\alpha$ ratios in two situations, e.g. in case of a
lognormal PDF: i) the lognormal dispersion $\sigma$ is much higher
than $\sigma\simeq2$, ii) the distance from the ionizing source is large (several hundred
kpc). In the likely hypothesis that a radio-galaxy is an AGN
with ionisation cones mostly oriented on the plane of the sky, the
projected distance will be similar to the actual separation.
We would  therefore expect to see a decrease in the $\ion{He}{ii}$/Ly$\alpha$
ratio moving away from the radio-galaxy, as effectively observed 
in some cases (e.g., \citealt{morais17}). Moreover, if local turbulence
is responsible for a broadening of the density PDF, we would also
expect that the $\ion{He}{ii}$/Ly$\alpha$ ratio should decrease where the
gas kinematics is broader. Although there are no studies to our 
knowledge in the literature that have directly addressed such 
a possible correlation, we do notice that some recent IFU 
observations of radio-galaxy haloes seem
to show that lower $\ion{He}{ii}$/Ly$\alpha$ ratios (and brighter Ly$\alpha$ emission)
correspond to regions with larger FWHM of non-resonant lines such as $\ion{He}{ii}$ 
 {
(see e.g., Figures 4 and 6 in \citealt{silva18})
}
If this simple picture is correct, gas clouds in radio-galaxy haloes 
should have lower $\ion{He}{ii}$/Ly$\alpha$ ratio than gas clouds in 
radio-quiet systems (that are intrinsically narrower in terms of kinematics) 
at a fixed three-dimensional distance and AGN luminosity.
If this is not observed, 
then some gas illuminated by type-I quasar must be intrinsically
more distant than in the radio-loud case. We have argued that
this is indeed the case for the ``bright tail'' in the Slug nebula.

This picture, from a geometrical perspective, could be seen as equivalent
to claiming that the ``observed'' illuminated volume of a quasar 
(or type-I AGN) should be much larger than the one of a radio-galaxy 
(or type-II AGN). This is indeed the case, when light travel effects
are considered for bright sources with ages less than a few Myr
(see e.g., \citealt{sc08} for an example): the size of the 
illuminated region along our line of sight, between us and the quasar,
is much larger than the size in the plane of the sky, or behind the source.
Especially if the AGN opening angle is small, observing 
around a type-I AGN would imply a higher probability of detecting
an object along our line of sight, if present, rather than in the transverse
direction. 

In our picture, radio-loud nebulae should be easier to detect 
because the $\sigma$ of their density distribution is increased
by interaction with the radio-jets (or other processes related to 
feedback) and therefore the Ly$\alpha$ SB will be increased
because of the elevated clumping factor.
Indeed radio-loud quasars at $z\simeq2.5$ do seem to have statistically 
brighter Ly$\alpha$ nebulae with respect to radio-quiet quasars 
as demonstrated already a few decades ago
(e.g., \citealt{heckman91}; see \citealt{sc17} for a review).
In general, the higher $\sigma$ could be however compensated by a smaller distance
with respect to the ``bright tail'' in the Slug nebula (or by smaller UV luminosity of the
AGN), producing higher $\ion{He}{ii}$/Ly$\alpha$ ratios in radio-galaxies with respect
to our case.  
The same would also apply to radio-quiet nebulae around type-II AGN, 
as long as they are kinematically broad because of local turbulence
or any other process that enhances $\sigma$.
The MAMMOTH-1 nebula around a radio-quiet type-II AGN \citep{cai17}
is one of such example: its Ly$\alpha$ SB are very high but $\ion{He}{ii}$
extended emission (with $\ion{He}{ii}$/Ly$\alpha\sim0.1$) is only confined
within 
 {30}
 kpc from the AGN (see also \citealt{prescott15} 
for another example).

If our interpretation is correct, Ly$\alpha$ nebulae as bright 
as the Slug nebula around type-I AGN should be therefore relatively rare
(as is indeed the case, see e.g., \citealt{sc17} for a review) because they represent 
the chance alignment between a large filament - containing haloes massive enough to
 produce a large $\sigma$ or large densities - a hyper-luminous quasar and 
 our line of sight. It is interesting to note that among the 100+ ubiquitous nebulae
around quasars at $z>3$ detected by MUSE in short exposure times
(e.g., \citealt{borisova16,fumagalli17, ginolfi18}; 
Arrigoni-Battaia et al., in prep.) only a handful show regions with Ly$\alpha$
SB as high as the Slug (once corrected for redshift dimming) 
at the projected distances of the Slug's ``bright tail'': MUSE Quasar nebulae \#1 and \#3
(MQN01 and MQN03) of \citet{borisova16}, and J1024+1040 of \citet{fab18}.
In all these cases there is no $\ion{He}{ii}$ detected in these distant regions 
with $\ion{He}{ii}$/Ly$\alpha<0.02$ (at 2$\sigma$ level). In particular, deep observations of 
J1024+1040 \citet{fab18} show many of the characteristic 
features discussed here: i) a region with high Ly$\alpha$
SB 
 {
($1.8\times10^{-17}$ erg s$^{-1}$ cm$^{-2}$ arcsec$^{-2}$)
}
at projected distance larger than 10 arcsec from the main quasar, ii)
low $\ion{He}{ii}$/Ly$\alpha$ ($<0.02$ at 2$\sigma$ level) in this region, iii) a large velocity shift 
 {
($\simeq300$ km/s)
}
,
iv) multiple quasars at different velocities but close projected separations. 
As in the case of the Slug, all these points hint at the possibility that
also J1024+1040 could be a large filament oriented along our line of sight
with similar properties to the Slug in terms of gas density distribution
(see the discussion in \citealt{fab18} for a different hypothesis).
Results of deeper observations of the two other possible analogous cases
- MQN01 and MQN03 - will be presented in future works (Cantalupo et al, in prep.).
As in the case of the Slug, however, detection of a non-resonant hydrogen line,
such as H$\alpha$ (only possible from space for $z>3$) would be fundamental 
to exclude also in these cases the possibility that Ly$\alpha$ is enhanced by
scattering of photons produced in the quasar broad line region. 

\section{Summary}\label{Sum}

We have reported the results of a deep MUSE search for extended $\ion{He}{ii}$ and metal 
emission from the Slug nebula at $z\simeq2.3$, 
one of the largest and most luminous giant Ly$\alpha$ nebulae discovered 
to date \citep{sc14}.
With the help of a new package for data reduction and analysis of MUSE datacubes
(CubExtractor; Cantalupo, in prep) briefly summarized in sections \ref{sec_obs} and \ref{sec_res},
we were able to detect and extract faint and diffuse emission associated with 
$\ion{He}{ii}\lambda1640$ down to about $5\times10^{-19}$ erg s$^{-1}$ cm$^{-2}$ arcsec$^{-2}$
(corresponding to a 3$\sigma$ confidence level for a pseudo aperture of 1 arcsec$^2$ 
and spectral width of 3.75$\mathrm{\AA}$).
The overall extent of the emission approaches 12", i.e. about 100 physical kpc.
$\ion{C}{iv}$ and $\ion{C}{iii}$ extended emission lines are instead 
not detected at significant levels in the current data, 
except at the location of a continuum and Ly$\alpha$ bright source ``c'' (see Figs. 1 and 7).
This source 
 {
is located in correspondence of the brightest Surface Brightness peak
}
in the extended $\ion{He}{ii}$ emission (dubbed ``region c'').

By comparing the positions of the $\ion{He}{ii}$ and Ly$\alpha$ emission and the implied 
$\ion{He}{ii}$/Ly$\alpha$ ratio we found, surprisingly, that the brightest Ly$\alpha$
emitting region (dubbed the ``bright tail'') just south of ``region c'' 
does not have any detectable $\ion{He}{ii}$ emission. 
The implied 2$\sigma$ upper limit for the $\ion{He}{ii}$/Ly$\alpha$ ratio in the
``bright tail'' is about $1\%$, significantly lower than some parts of
``region c'' that are located just a few arcsec to the north 
and which reach $\ion{He}{ii}$/Ly$\alpha$ ratios up to $8\%$ 
(the typical $\ion{He}{ii}$/Ly$\alpha$ ration in the ``region c'' is about $5\%$, see Fig. 6).

We discussed the possible origins for such a strong apparent gradient
in the observed $\ion{He}{ii}$/Ly$\alpha$ ratio in section \ref{Disc} concluding that 
the most likely scenario requires  
that the ``bright tail'' must be located at a physical distance from quasar ``a''
that is about three times larger than the distance between quasar ``a'' and
``region c''. This is corroborated by the similarity between the 
 {$\ion{He}{ii}$/Ly$\alpha$}
line ratio gradient
and velocity shifts of both Ly$\alpha$ and H$\alpha$ emission 
(presented in \citealt{leibler18}) associated with the same regions. 

We then examined which physical situations would be able to produce such
a low $\ion{He}{ii}$/Ly$\alpha$ ratios and argued that the ``observed'' line ratios
(i.e., the ratio of observed line fluxes measured through an aperture)
of recombination lines will be lower than the ``intrinsic'' values 
if the emitting gas density is not constant, 
i.e. if the gas distribution is not a delta function as typically assumed in the literature
(section \ref{Disc_sub1}).
By assuming a log-normal density distribution and that the cold emitting
gas is on average in pressure equilibrium with a confining hot gas,
we found (see Fig. \ref{Fig_LN_LR}) that a log-normal dispersion
$\sigma\sim2.5$ and volume-averaged densities of  $\langle n \rangle_c\sim0.05$
cm$^{-3}$ for the cold component are able to explain the line
ratios of both ``region c'' and the ``bright tail'' if they are placed
 {
at a distance of about 270 kpc and at least 900 kpc, 
respectively, from quasar ``a''
(assuming Case B recombination; for Case A, these distances
reduce to about 180 kpc and at least 550 kpc for ``region c" and 
the ``bright tail", respectively).
}
We noted that such high $\sigma$ are not
dissimilar to what is expected in the interstellar medium of
galaxies.

Our results imply - on the large scales - that the Slug nebula 
 {
could be composed by multiple structures as a
}
 part of a large filamentary structure
extending on scales of about a physical Mpc 
and mostly oriented along our line of sight. 
On the other hand, our analysis also confirms that on small scales (below
our current resolution limit of a few kpc) the gas density distribution
in such structures should be extremely broad or clumpy,
possibly indicating a very turbulent medium. 
Finally, by putting our results in the
broader context of Ly$\alpha$ nebulae discovered 
around other quasars and type-II AGN such as radio galaxies, 
we argued that both geometrical and density-distribution
effects are fundamental drivers 
of both Ly$\alpha$ and $\ion{He}{ii}$ surface brightnesses  
among these systems.

\section*{Acknowledgements}

SC and GP gratefully acknowledge support from Swiss National Science Foundation grant PP00P2\_163824.
RB and AF acknowledges support from the ERC advanced grant 339659-MUSICOS.
Part of the figures in this manuscript have been realised with the open source
software VisIt \citep{VisIt} supported by the Department of Energy with funding from the 
Advanced Simulation and Computing Program and the Scientific Discovery 
through Advanced Computing Program.

\bibliographystyle{mnras}
\bibliography{MUSESlug_arXiv}

\appendix

\section{comparison between Case A and Case B line ratios}

\begin{figure}
\includegraphics[width=1.0\columnwidth]{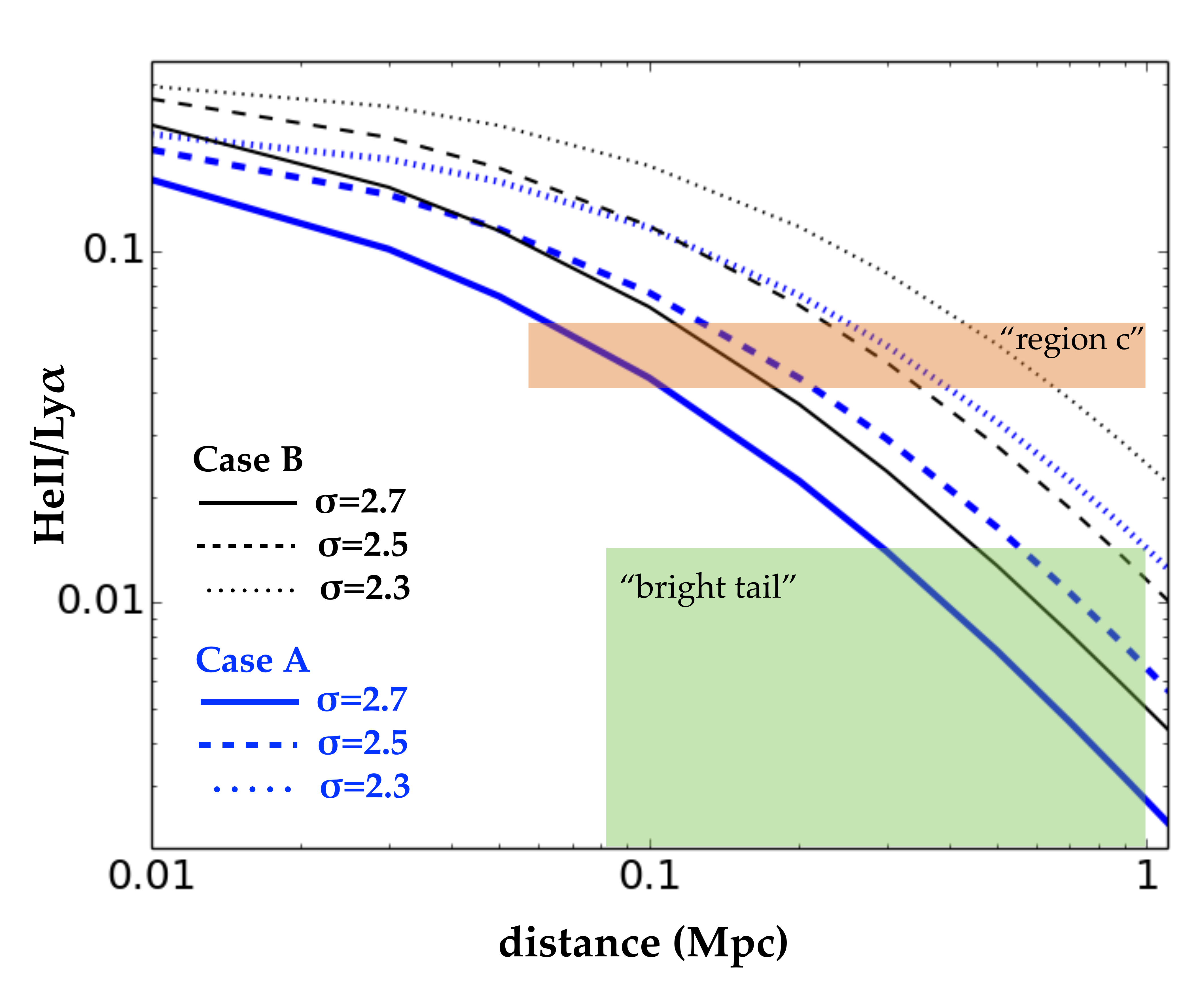}
\caption{observed $\ion{He}{ii}$/Ly$\alpha$ ratio obtained by our photoionisation
model using a log-normal density distribution with $\langle n \rangle_c=0.05$ cm$^{-3}$
for both Case A and Case B effective and total recombination coefficients.
The results are shown as a function of distance from the ionizing
source for three values of the lognormal dispersion $\sigma$
(see legend for details). The shaded areas represent the allowed parameter space 
for both the ``region c" and the ``bright tail" regions as described
in detail in the text.
}\label{FigLRvsDist}
\end{figure}

 {
Here we present the results of our line ratio calculations using both 
Case A and Case B effective and total recombination coefficients
for a log-normal density distribution using the model described in
section \ref{subsubLN}.
In particular, in Fig.\ref{FigLRvsDist}, we show 
the $\ion{He}{ii}$/Ly$\alpha$ ratios obtained by fixing the average 
$\langle n \rangle_c$ to $0.05$ cm$^{-3}$ and for three values
of the lognormal dispersion $\sigma$
as a function of the distance from the ionising source
(see legend in figure for details).
The shaded areas represent the allowed parameter space 
for both the ``region c" and the ``bright tail" region considering
the measured $\ion{He}{ii}$/Ly$\alpha$ ratio including both noise and systematic errors
and the 2$\sigma$ noise upper limit, respectively. 
The minimum and maximum distance are chosen from 
the observed projected distance and the distance 
at which the region becomes self-shielded to the
hydrogen ionizing radiation, respectively. 
The line ratios obtained using the Case A coefficients
are always smaller by about a factor 1.7 at a given
$\sigma$ and distance in the range of plausible
distances for both ``region c" and the ``bright tail". 
This translate in a smaller required distance by a similar factor 
for both regions in order to obtain the same line ratios as in the Case B,
as discussed in detail in section \ref{subsubLN}.
}

\bsp    
\label{lastpage}
\end{document}